\documentclass[journal]{IEEEtran}

\usepackage{amsmath,amssymb,amsthm}
\usepackage{cite}
\usepackage{epsfig}
\usepackage[ruled,vlined,lined,linesnumbered]{algorithm2e}
\usepackage{color}
\usepackage{threeparttable}
\ifCLASSINFOpdf
\else
\fi
\newcommand{\defeq}{\stackrel{\text{def}}{=}}
\theoremstyle{definition}
\newtheorem{PROP}{Proposition}

\begin{document}
%
% paper title
% can use linebreaks \\ within to get better formatting as desired
\title{Symbol-Decision Successive Cancellation List Decoder for Polar Codes}

% author names and affiliations
% use a multiple column layout for up to three different
% affiliations
\IEEEoverridecommandlockouts

\author{
\IEEEauthorblockN{Chenrong Xiong, Jun Lin~\IEEEmembership{Student~member,~IEEE} and Zhiyuan Yan,~\IEEEmembership{Senior~member,~IEEE}}
%\thanks{This work was supported in part by NSF under Grants ECCS-0925890 and ECCS-1055877.}

%\IEEEauthorblockA{Department of Electrical and Computer Engineering\\
%Lehigh University\\
%Bethlehem, PA 18015 USA}

%\IEEEauthorblockN{Michael Shell}
%\IEEEauthorblockA{School of Electrical and\\Computer Engineering\\
%Georgia Institute of Technology\\
%Atlanta, Georgia 30332--0250\\
%Email: http://www.michaelshell.org/contact.html}
%\and
%\IEEEauthorblockN{Homer Simpson}
%\IEEEauthorblockA{Twentieth Century Fox\\
%Springfield, USA\\
%Email: homer@thesimpsons.com}
%\and
%\IEEEauthorblockN{James Kirk\\ and Montgomery Scott}
%\IEEEauthorblockA{Starfleet Academy\\
%San Francisco, California 96678-2391\\
%Telephone: (800) 555--1212\\
%Fax: (888) 555--1212}
}

% conference papers do not typically use \thanks and this command
% is locked out in conference mode. If really needed, such as for
% the acknowledgment of grants, issue a \IEEEoverridecommandlockouts
% after \documentclass

% for over three affiliations, or if they all won't fit within the width
% of the page, use this alternative format:
%
%\author{\IEEEauthorblockN{Michael Shell\IEEEauthorrefmark{1},
%Homer Simpson\IEEEauthorrefmark{2},
%James Kirk\IEEEauthorrefmark{3},
%Montgomery Scott\IEEEauthorrefmark{3} and
%Eldon Tyrell\IEEEauthorrefmark{4}}
%\IEEEauthorblockA{\IEEEauthorrefmark{1}School of Electrical and Computer Engineering\\
%Georgia Institute of Technology,
%Atlanta, Georgia 30332--0250\\ Email: see http://www.michaelshell.org/contact.html}
%\IEEEauthorblockA{\IEEEauthorrefmark{2}Twentieth Century Fox, Springfield, USA\\
%Email: homer@thesimpsons.com}
%\IEEEauthorblockA{\IEEEauthorrefmark{3}Starfleet Academy, San Francisco, California 96678-2391\\
%Telephone: (800) 555--1212, Fax: (888) 555--1212}
%\IEEEauthorblockA{\IEEEauthorrefmark{4}Tyrell Inc., 123 Replicant Street, Los Angeles, California 90210--4321}}

% use for special paper notices
%\IEEEspecialpapernotice{(Invited Paper)}

% make the title area
\maketitle
\begin{abstract}
Polar codes are of great interests because they provably achieve the
capacity of both discrete and continuous memoryless channels
while having an explicit construction. Most existing decoding algorithms of polar codes
are  based on bit-wise hard or soft decisions. In this paper, we propose
symbol-decision successive cancellation (SC) and successive cancellation list (SCL) decoders for polar codes, which use symbol-wise
hard or soft decisions for higher throughput or better error performance. First, we propose to use a
recursive channel combination to calculate symbol-wise channel
transition probabilities, which lead to symbol decisions. Our proposed recursive channel combination also has a lower complexity than
simply combining bit-wise channel
transition probabilities. The similarity between our proposed method and Arikan's channel
transformations also helps to share hardware resources between calculating bit- and
symbol-wise channel transition probabilities.
Second, a two-stage list pruning network is proposed to provide a trade-off between the error performance
and the complexity of the symbol-decision SCL decoder. Third, since memory is a significant part of SCL decoders, we propose a pre-computation memory-saving
technique to reduce memory requirement of an SCL decoder. Finally, to evaluate the throughput advantage of our symbol-decision decoders, we design an
architecture based on a semi-parallel successive cancellation list decoder. In this architecture,
different symbol sizes, sorting implementations, and message scheduling schemes are considered.
%Two-, four- and eight-bit symbol-decision SCL decoders are implemented.
Our synthesis results show that in terms of area efficiency, our symbol-decision SCL decoders outperform both bit- and symbol-decision SCL
decoders.
\end{abstract}


% For peer review papers, you can put extra information on the cover
% page as needed:
% \ifCLASSOPTIONpeerreview
% \begin{center} \bfseries EDICS Category: 3-BBND \end{center}
% \fi
%
% For peerreview papers, this IEEEtran command inserts a page break and
% creates the second title. It will be ignored for other modes.
\IEEEpeerreviewmaketitle
\section{Introduction}
\label{sec:intro}
Polar codes, a groundbreaking finding by Arikan \cite{5075875} in 2009, have
ignited a spark of research interest in the fields of communication
and coding theory, because they can provably achieve the capacity for both
discrete \cite{5075875} and continuous \cite{5351487}
memoryless channels. The second reason why polar codes are
attractive is their low encoding and decoding complexity. For example, a polar code of length $N$ can be decoded by
the successive cancellation (SC) algorithm \cite{5075875} with a
complexity of $\mathcal{O}(N\log N)$. However, their capacity
approaching can be achieved only when the code length is large enough ($N >
2^{20}$ \cite{6327689}) if the SC algorithm is used. For short or
moderate code length, in terms of the error performance,
polar codes with the SC algorithm are inferior to Turbo codes or low-density
parity-check (LDPC) codes \cite{6297420, 6033837}.

Since the debut of polar codes, a lot of efforts have been made to improve the
error performance of short polar codes. Systematic polar codes \cite{5934670} were
proposed to reduce the bit error rate (BER) while guaranteeing the same frame
error rate (FER) as their non-systematic counterparts. Although a Viterbi
algorithm \cite{ML_polar}, a sphere decoding algorithm \cite{6283643} and stack
sphere decoding algorithm \cite{6708139} can provide maximum
likelihood (ML) decoding of polar codes, they are considered infeasible,
especially for long polar codes, due to their much higher complexity than the SC algorithm.
Recently, an SC list algorithm for polar codes was proposed in
\cite{6033904} to bridge the performance gap between the SC algorithm and ML algorithms at the cost of complexity of $\mathcal{O}(LN\log N)$,
where $L$ is the list size. Moreover, the concatenation of polar codes with
cyclic redundancy check (CRC) codes was introduced in \cite{6297420, Tal2012}. To decode
the CRC-concatenated polar codes, a CRC detector is used in the SCL
algorithm to help select the output codeword. The combination of an SCL
algorithm and a CRC detector is called CRC-aided SCL (CA-SCL)
algorithm. \cite{Tal2012} shows that with the CA-SCL algorithm, the
error performance of a (2048, 1024) CRC-concatenated polar code is better that
of a (2304, 1152) LDPC code, which is used in the WiMax standard \cite{1603394}.

Several architectures have been proposed for the SC algorithm. Arikan
\cite{5075875} showed that a fully parallel SC decoder has a latency of $2N-1$
clock cycles. A tree SC decoder and a line SC decoder with complexity
of  $\mathcal{O}(N)$ were proposed in \cite{5946819}. These two decoders have
the same latency as the fully parallel SC decoder. To reduce complexity
further, Leroux \emph{et~al}.\ \cite{6327689} proposed a semi-parallel SC decoder for polar
codes by taking advantage of the recursive structure of polar codes to reuse
processing resources. Assuming that the number of processing elements (PEs) are $P$
$(P=2^p\leq N)$, the latency of the semi-parallel SC decoder is
$2N+\frac{N}{P}\log_2 (\frac{N}{4P})$ clock cycles. To reduce the latency, a simplified SC (SSC) polar decoder
was introduced in \cite{6065237} and it was further analyzed in
\cite{6680761}. In the SSC polar decoder, a polar code is converted to a binary
tree including three types of nodes: rate-one, rate-zero and rate-$R$
nodes. Based on the SSC polar decoder, the ML SSC decoder
makes use of the ML algorithm to deal with part of
rate-$R$ nodes in \cite{6464502}. However, the SSC and  ML-SSC polar decoders
depend on positions of information bits and frozen bits, and are code-specific
consequently. In \cite{6475198}, a pre-computation look-ahead technique was
proposed to reduce the latency of the tree SC decoder by half. For the SCL polar
decoder, the semi-parallel architecture was adopted in \cite{6865312}. In \cite{6823099} Balatsoukas-Stimming \emph{et~al}.\ proposed an architecture of $L=4$ to achieve a
throughput of 124 Mbps and a latency of 8.25 ms when decoding a (1024, 512) polar code. In
\cite{JunPolarList}, Lin and Yan designed an SCL polar decoder with the throughput of 182
Mbps and a latency of 5.63 ms. To reduce the memory requirement, the
log-likelihood ratio (LLR) messages are used in \cite{LLR-SC-TSP}. The throughput of existing polar decoders is
still not high enough for high speed applications.

Since the low throughput (or long latency) of the SC decoder is due to its serial nature, several previous works attempt to improve the throughput (or latency).
In \cite{ParSC}, the data bits of a polar code is split into several streams, which are decoded simultaneously. This idea of parallel processing is extended in \cite{ParSC3}, where the SC decoder is transformed into a concatenated decoder, where all the inner SC decoders are carried out in parallel. Yuan and Parhi proposed a multi-bit SCL decoder \cite{ParSC2}.

%Recently, parallel SC decoder and SCL decoder of polar codes were proposed
%in\cite{ParSC} and \cite{ParSC2} to improve throughput by calculating the
%symbol-based channel transition probability and making decision for
%multiple bits at a time. Both \cite{ParSC} and \cite{ParSC2} use matrix
%operation to find a direct mapping between output partial sums and the encoding
%data, and then calculate symbol-based channel transition probabilities based on this mapping.
%
%In this paper, we propose a new method to calculate the symbol-based channel transition
%probability from a perspective of recursive channel combination. This method
%reveals the similarity of calculating the bit- and symbol-based channel
%transition probabilities. However, \cite{ParSC} and \cite{ParSC2} neglect this
%similarity and use additional dedicated adders to calculate the
%symbol-based channel transition probabilities. Henceforth, the new SC/SCL decoding
%algorithm using our new proposed method is called the symbol-decision SC/SCL
%decoding algorithm. In this sense, the symbol-decision decoders are
%generalization of bit-decision decoders, and bit-decision decoders are special
%cases of symbol size being \emph{one}. The performance of the symbol-decision
%SC/SCL algorithm should not be worse than the bit-decision SC/SCL algorithm
%\cite{PerAnaSDSC}. Moreover, we design efficient decoder architectures for
%symbol-decision SCL decoders.

In this paper, we address the throughput/latency issue by proposing symbol-decision SC and SCL decoders, which are based on symbol-wise hard or soft decisions. Since each symbol consists of $M$ bits, when $M>1$ the symbol-decision decoders achieve higher throughput as well as better error performance. The proposed symbol-decision decoders are natural generalization of their bit-wise counterparts, and reduce to existing bit-wise decoders when the symbol size is one bit. The main contributions of this paper are:
\begin{itemize}
\item We propose a novel recursive channel combination to calculate the symbol-wise channel
  transition probabilities, which enable symbol decisions in SC and SCL algorithms.
  The proposed recursive channel combination also has a lower complexity than simply combining bit-wise channel transition probabilities.
  %  This   recursive scheme is derived based on the Arikan's recursive channel transformation introduced by Arikan in \cite{5075875}.
%  Our proposed method needs fewer additions than ML decoders
%  used in \cite{ParSC} and \cite{ParSC2}.
The similarity between the Arikan's recursive channel transformation
  and our symbol-wise recursive channel combination helps to share hardware resources to
  calculate the bit- and symbol-based channel transition probabilities.
\item An $M$-bit symbol-decision SCL decoder needs to find the $L$ most reliable
  candidates out of $2^ML$ list candidates. We propose a two-stage list pruning
  network to perform this sorting function. This pruning network also provides a
  trade-off between performance and complexity.
\item By adopting pre-computation technique \cite{135746}, We develop a
  pre-computation memory-saving (PCMS) technique to reduce the memory
  requirement of the SCL decoder. Specifically, the channel information memory
  can be eliminated when using the PCMS technique. Moreover, this
  technique also helps to improve throughput slightly.
\item To evaluate the throughput of symbol-decision SC decoders, we propose an area efficient architecture for symbol-decision SCL
  decoders\footnote{We focus on the SCL decoder because the SC decoder can be considered as an SCL
  decoder with a list size of one.}. In our architecture, to save the area, adders in processing units
  are reused to calculate the symbol-wise channel transition probability. We propose two scheduling schemes for  sharing hardware resources. We also propose
  two list pruning network for designs with different symbol sizes. 
\item   We design two-, four-, and eight-bit
symbol-decision SCL decoders for a (1024, 480) CRC32-concatenated polar code with a list size of
four. Synthesis results show that in terms of area efficiency, our symbol-decision SCL
decoder outperforms all existing state-of-the-arts SCL decoders in \cite{ParSC2,
  JunPolarList, 6823099, LLR-SC-TSP}. For example, the area
efficiency of our four-bit symbol-decision SCL decoder is 259.2
Mb/s/$\text{mm}^2$, which is 1.51 times as big as that of
\cite{LLR-SC-TSP}. Our implementation results also demonstrate that the
symbol-decision SCL decoder can provide a range of tradeoffs between area,
throughput, and area efficiency.
\end{itemize}

Our symbol-decision decoding algorithms assume that the underlying channel has a binary input, and our symbol-wise channel transformation is virtual and introduced for decoding only. Hence, our work is different from those assuming a $q$-ary ($q>2$) channel (see, for example, \cite{Park_Barg_IT_Feb13}). 

The decoding schedule (bit sequence) of our symbol-decision decoding algorithms is actually the same as those in \cite{ParSC,ParSC2,ParSC3}, but our symbol-decision decoding algorithms are different from those in \cite{ParSC,ParSC2,ParSC3} in two aspects. First, our symbol-wise recursive channel transition is different from how transition probabilities are derived in \cite{ParSC,ParSC2,ParSC3}. Second, the symbol-decision perspective allows us to prove that the symbol-decision algorithms have better frame error rates (FERs) than their bit-decision counterparts \cite{FER_performance}, while only simulation results are provided in \cite{ParSC,ParSC2} and error performance is not investigated in \cite{ParSC3}. There are additional differences between our decoding algorithms/architectures and those in \cite{ParSC,ParSC2,ParSC3}. For instance, all the bits within a symbol are estimated jointly in our symbol-decision SC algorithm, whereas some bits are decoded independently for the decoder with parallelism two in \cite{ParSC}. Also, while our symbol-decision decoding is introduced on the algorithmic level, the multibit decoder is introduced on the level of decoding operations \cite{ParSC2}. Finally, for our symbol-decision SCL decoders, we use the semi-parallel architecture
  because it is more area efficient than the tree architecture and the line
  architecture \cite{5946819}. %In contrast, the multibit decoder architecture in \cite{ParSC2} is based on ???.

The rest of our paper is organized as follows. Section~\ref{sec:review} briefly reviews
 polar codes and existing decoding algorithms for polar codes. In
 Section~\ref{sec:SBDecoder}, the symbol-based
recursive channel combination is proposed to calculate the symbol-based
channel transition probability. Moreover, to simplify the selection of
the list candidates, a two-stage list pruning network
is proposed. In Section~\ref{sec:PCMS}, we
introduce a method to reduce memory requirement of list decoders of polar
codes by pre-computation technique. In Section~\ref{sec:ImpSBDecoder}, we demonstrate the hardware architecture for symbol-decision SCL decoders. Two scheduling
schemes for hardware sharing are discussed. We also propose two list pruning
network for different designs: a folded sorting implementation and a tree sorting
implementation. A discussion on the latency of our architecture and synthesis results for our
implementations are provided in this section as well. Finally, we draw some conclusions in Section~\ref{sec:conclusion}.

\section{Polar Codes and Existing Decoding Algorithms}
\label{sec:review}
\subsection{Preliminaries}
We follow the notation for vectors in \cite{5075875}, namely
$u_a^b=(u_a,u_{a+1},\cdots,u_{b-1},u_b)$; if $a>b$, $u_a^b$ is regarded as
void. $u_{a,o}^{b}$ and $u_{a,e}^{b}$ denote the subvector of $u_a^b$ with odd
and even indices, respectively.

Let $W:\mathcal{X}\rightarrow\mathcal{Y}$ represent a generic B-DMC
with binary input alphabet $\mathcal{X}$, arbitrary output alphabet $\mathcal{Y}$, and transition
probabilities $W(y|x)$, $y\in \mathcal{Y}$, $x\in\{0,1\}$. Assume $N$ is an
arbitrary integer and $M$ is an integer satisfying $M\vert N$. Let $W_{N,M}^{(j)}$s
denote a set of $\frac{N}{M}$ coordinate channels: $W_{N,M}^{(j)}:
\mathcal{X}^M\rightarrow\mathcal{Y}^N\times\mathcal{X}^{(j-1)M}$, $0 < j \leq \frac{N}{M}$ with
the transition probabilities $W_{N,M}^{(j)}(y_1^N,x_1^{(j-1)M}|x_{(j-1)M+1}^{jM})$, where
$(y_1^N,x_1^{(j-1)M})$ and $x_{(j-1)M+1}^{jM}$ denote the output and input of
$W_{N,M}^{(j)}$, respectively.

\subsection{Polar Codes}

Polar codes are linear block codes, and their block lengths are
restricted to powers of two, denoted by $N=2^n$ for $n \geq 2$.  Assume $\mathbf{u}=u_1^N=(u_1,u_2,\cdots,u_{N})$ is the data bit
sequence. Let 
$F=\left[
\begin{smallmatrix}
1 & 0 \\
1 & 1 
\end{smallmatrix}
\right]$. The corresponding encoded bit sequence
$\mathbf{x}=x_1^{N}=(x_1,x_2,\cdots,x_{N})$ is generated by
\begin{equation}
\mathbf{x} = \mathbf{u}B_NF^{\otimes n},
\label{equ:encoder}
\end{equation}
where $B_N$ is the $N\times N$ bit-reversal permutation matrix and $F^{\otimes
  n}$ denotes the $n$-th Kronecker power of $F$ \cite{5075875}. 

For any index set $\mathcal{A} \subseteq \{1,2,\cdots, N\}$, $\mathbf{u}_{\mathcal{A}}=(u_i: 0 < i \leq N, i\in
\mathcal{A})$ is the sub-sequence of $\mathbf{u}$ restricted to $\mathcal{A}$. For an $(N,K)$ polar code, the data bit sequence is
grouped into two parts: a $K$-element part $\mathbf{u}_{\mathcal{A}}$ which
carries information bits, and $\mathbf{u}_{\mathcal{A}^c}$ whose elements are
predefined frozen bits, where $\mathcal{A}^c$ is the complement of $\mathcal{A}$. For convenience, frozen bits are set to zero. 

\subsection{SC Algorithm for Polar Codes}
Given a transmitted codeword $\mathbf{x}$ and the corresponding received word
$\mathbf{y}$, the SC algorithm for an $(N,K)$ polar code estimates the
encoding bit sequence $\mathbf{u}$ successively as shown in Alg.~\ref{alg:SC}.
Here, $\hat{\mathbf{u}} = (\hat{u}_1,
\hat{u}_2,\cdots,\hat{u}_N)$  represents the estimated value for
$\mathbf{u}$.
%$\Pr(\mathbf{y},\hat{u}_1^{j-1}|u_j)$ is the
%probability that $\mathbf{y}$ is received and the previously decoded bits are $\hat{u}_1^{j-1}$
%given $u_j$ is zero or one. 

\begin{algorithm}
\caption{SC Decoding Algorithm \cite{5075875}}
\label{alg:SC}
\LinesNumbered
\For{$j=1:N$}{
\lIf{$j\in \mathcal{A}^c$}{
$\hat{u}_j=0$
}
\uElse{
\lIf{$\frac{W^{(j)}_{N,1}(\mathbf{y},\hat{u}_1^{j-1}|u_j=1)}{W^{(j)}_{N,1}(\mathbf{y},\hat{u}_1^{j-1}|u_j=0)} \geq 1 $}{
$\hat{u}_j = 1$
}
\lElse{
$\hat{u}_j = 0$
}
}
}
\end{algorithm}

To calculate $W_{N,1}^{(j)}(\mathbf{y},\hat{u}_1^{j-1}|u_j)$, Arikan's recursive
channel transformation \cite{5075875} is applied. A pair of binary channels
$W_{2\Lambda,1}^{(2i-1)}$ and $W^{(2i)}_{2\Lambda,1}$ are obtained by a
single-step transformation of two independent copies of a binary input channel
$W^{(i)}_{\Lambda,1}$ : $(W^{(i)}_{\Lambda,1},W^{(i)}_{\Lambda,1})
\mapsto (W_{2\Lambda,1}^{(2i-1)},W^{(2i)}_{2\Lambda,1})$. The channel transition
probabilities of $W_{2\Lambda,1}^{(2i-1)}$ and $W^{(2i)}_{2\Lambda,1}$ are
given by

\begin{equation}
\begin{split}
\label{eq:ArikanT1}
%\Pr&(y_1^{N},u_1^{2i}|u_{2i+1})\\
%&=\frac{1}{2}\sum_{u_{2i+2}}\Bigl[\Pr(y_{1}^{{N}/2},u_{1,o}^{2i}\oplus u_{1,e}^{2i}|u_{2i+1}\oplus
%u_{2i+2})\\
%&\hspace{20mm}\cdot\Pr(y_{{N}/2+1}^{{N}},u_{1,e}^{2i}|u_{2i+2})\Bigr],
W&_{2\Lambda,1}^{(2i-1)}(y_1^{2\Lambda},u_1^{2i-2}|u_{2i-1})\\
&=\frac{1}{2}\sum_{u_{2i}}\Bigl[W^{(i)}_{\Lambda,1}(y_{1}^{\Lambda},u_{1,o}^{2i-2}\oplus u_{1,e}^{2i-2}|u_{2i-1}\oplus
u_{2i})\\
&\hspace{20mm}\cdot W^{(i)}_{\Lambda,1}(y_{\Lambda+1}^{2\Lambda},u_{1,e}^{2i-2}|u_{2i})\Bigr],
\end{split}
\end{equation}
and
\begin{equation}
\begin{split}
\label{eq:ArikanT2}
%\Pr&(y_1^{N},u_1^{2i+1}|u_{2i+2})\\
%&=\frac{1}{2}\Pr(y_{1}^{{N}/2},u_{1,o}^{2i}\oplus u_{1,e}^{2i}|u_{2i+1}\oplus
%u_{2i+2})\\
%&\hspace{20mm}\cdot\Pr(y_{{N}/2+1}^{{N}},u_{1,e}^{2i}|u_{2i+2}),
W&^{(2i)}_{2\Lambda,1}(y_1^{2\Lambda},u_1^{2i-1}|u_{2i})\\
&=\frac{1}{2}W^{(i)}_{\Lambda,1}(y_{1}^{\Lambda},u_{1,o}^{2i-2}\oplus u_{1,e}^{2i-2}|u_{2i-1}\oplus
u_{2i})\\
&\hspace{20mm}\cdot W^{(i)}_{\Lambda,1}(y_{\Lambda+1}^{2\Lambda},u_{1,e}^{2i-2}|u_{2i}),
\end{split}
\end{equation}
where $0 < i \leq \Lambda=2^{\lambda} < N$, and $0
\leq \lambda < n$.

Expressed in log-likelihood (LL), Eqs.~\eqref{eq:ArikanT1} and \eqref{eq:ArikanT2} can
be approximated as \cite{6297420}:
\begin{equation}
\label{eq:LL_T_1}
\begin{split}
{\rm LL}&_{2\Lambda}^{(2i-1)}(y_1^{2\Lambda},u_1^{2i-2}|u_{2i-1})\\
& \approx \max\biggl\{\Bigl[{\rm
  LL}_{\Lambda}^{(i)}(y_{1}^{\Lambda},u_{1,o}^{2i-2}\oplus u_{1,e}^{2i-2}|u_{2i-1}\oplus
0)\\
&\hspace{20mm}+{\rm LL}_{\Lambda}^{(i)}(y_{\Lambda+1}^{2\Lambda},u_{1,e}^{2i-2}|0)\Bigr],\\
&\hspace{13mm}\Bigl[{\rm
  LL}_{\Lambda}^{(i)}(y_{1}^{\Lambda},u_{1,o}^{2i-2}\oplus u_{1,e}^{2i-2}|u_{2i-1}\oplus
1)\\
&\hspace{20mm}+{\rm
  LL}_{\Lambda}^{(i)}(y_{\Lambda+1}^{2\Lambda},u_{1,e}^{2i-2}|1)\Bigr]\biggr\}-\log 2,
\end{split}
\end{equation}
\begin{equation}
\label{eq:LL_T_2}
\begin{split}
{\rm
  LL}&_{2\Lambda}^{(2i)}(y_1^{\Lambda},u_1^{2i-1}|u_{2i})\\
& \approx {\rm
  LL}_{\Lambda}^{(i)}(y_{1}^{\Lambda},u_{1,o}^{2i-2}\oplus u_{1,e}^{2i-2}|u_{2i-1}\oplus
u_{2i})\\
&\hspace{20mm}+{\rm LL}_{\Lambda}^{(i)}(y_{\Lambda+1}^{2\Lambda},u_{1,e}^{2i-2}|u_{2i})-
\log 2.
\end{split}
\end{equation}

To simplify the calculation, the constants in
Eqs.~\eqref{eq:LL_T_1} and \eqref{eq:LL_T_2} can be discarded since this global offset for all
LLs does not affect the decoding decision.

\subsection{Parallel SC Algorithm for Polar Codes}
\begin{figure}[htbp]
\centering
\includegraphics[width=8cm]{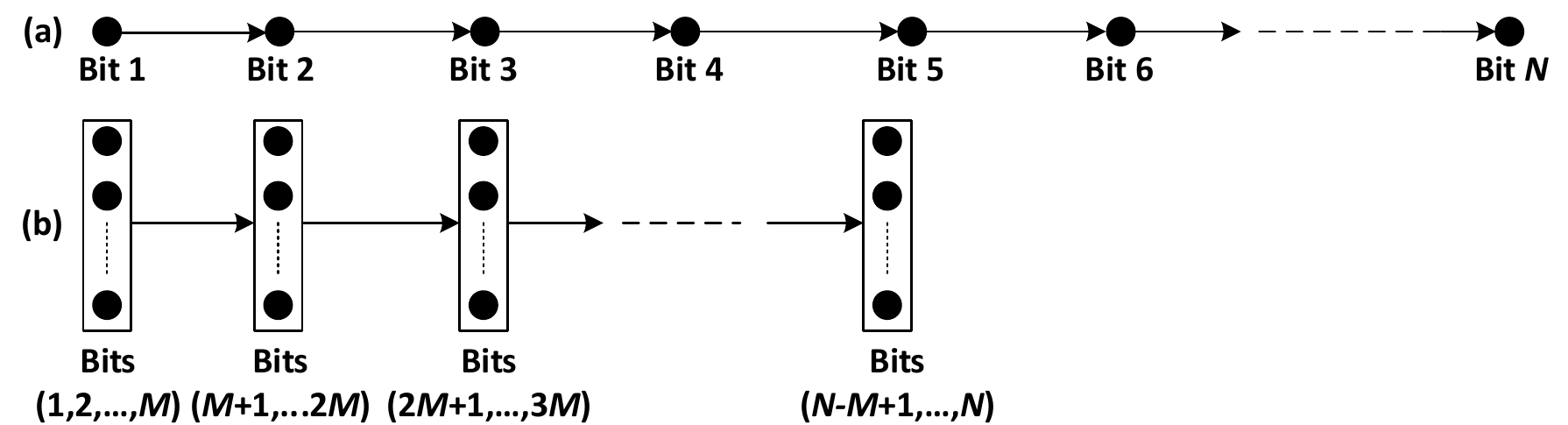}
\caption{Decoding of (a) bit-decision vs. (b) $M$-bit symbol-decision}
\label{fig:BDSD}
\end{figure}

The SC algorithm makes hard-decision for only one bit at a
time, as shown in Fig.~\ref{fig:BDSD}(a). We call it bit-decision decoding
algorithm. A parallel SC decoder \cite{ParSC, ParSC2, ParSC3} makes
hard-decision for $M$ bits instead of only one bit at a time, as shown in
Fig.~\ref{fig:BDSD}(b).

Without loss of generality, assume $M$ is a
power of two, i.e.\ $M=2^m( 0\leq m \leq n)$. $\mathcal{IM}_j \defeq
\{jM-M+1,jM-M+2,\cdots,jM\}$, for $0 < j \leq \frac{N}{M}$. $\mathcal{AM}_j \defeq \mathcal{IM}_j \cap \mathcal{A}\hspace{3mm}{\rm and}\hspace{3mm} \mathcal{AM}_j^c \defeq \mathcal{IM}_j \cap \mathcal{A}^c$.

Given $\mathbf{y}$ and $\hat{u}_1^{jM-M}$, $u_{jM-M+1}^{jM}$ is determined by
%in Eq.~\eqref{equ:MB_DR}.
\begin{equation}
\label{equ:MB_DR}
\hat{u}_{jM-M+1}^{jM} =
\underset{\substack{
u_{\mathcal{AM}_j}\in\{0,1\}^{\lvert\mathcal{AM}_j\rvert} \\
u_{\mathcal{AM}_j^c}\in\{0\}^{\lvert\mathcal{AM}_j^c\rvert}}}{\arg\max}W_{N,M}^{(j)}(\mathbf{y},\hat{u}_1^{jM-M}|u_{jM-M+1}^{jM}),
\end{equation}
where $\lvert\mathcal{AM}_j\rvert$ represents the cardinality of
$\mathcal{AM}_j$. If $M=N$, this
decoding algorithm is exactly a maximum-likelihood sequence decoding algorithm.
%An $M$-bit symbol-decision polar decoder needs $M$ component decoders. Based on outputs of component
%decoders, an $M$-bit symbol-decision polar decoder decodes $M$ bits at a time instead of
%only one bit. However, it claimed that an $M$-bit symbol-decision polar decoder has $M$ tximes faster decoding speed
%than its bit-decision counterpart. For example, by
%using the fully parallel SC decoder as the component decoder, an $M$-bit
%symbol-decision polar decoder have complexity of $\mathcal{O}(N \log (N/M))$ and a
%decoding latency of $2\frac{N}{M}$ in an ideal case, under the assumption of
%that $M$ bits can be decoded in one clock cycle after receiving outputs of
%component decoders. This assumption is impractical for real applications,
%specifically, for high speed applications. Hardware complexity and computation
%delay should be taken into account.

\subsection{SCL and CA-SCL  Algorithms for Polar Codes}

\begin{algorithm}
\caption{SCL Decoding Algorithm \cite{6033904}}
\label{alg:SCL}
\LinesNumbered
$\alpha=1$\;
\For{$j=1:N$}{
\uIf{$j\in \mathcal{A}^c$}{
\For{$i=1:\alpha$}{
$(\mathcal{L}_i)_j = 0$\;
}
}
\uElseIf{$2\alpha\leq L$}{
\For{$i=1:\alpha$}{
$(\mathcal{L}_i)_1^j=${\tt conc}$((\mathcal{L}_i)_1^{j-1},0)$\;
$(\mathcal{L}_{i+\alpha})_1^j=${\tt conc}$((\mathcal{L}_i)_1^{j-1},1)$\;
}
$\alpha=2\alpha$\;
}
\Else{
\For{$i=1:L$}{
${\sf S}[i]{\sf .P}=W_{N,1}^{(j)}(\mathbf{y},(\mathcal{L}_i)_1^{j-1}|0)$\;
${\sf S}[i]{\sf .L}=(\mathcal{L}_i)_1^{j-1}$\;
${\sf S}[i]{\sf .U}=0$\;
${\sf S}[i+L]{\sf .P}=W_{N,1}^{(j)}(\mathbf{y},(\mathcal{L}_i)_1^{j-1}|1)$\;
${\sf S}[i+L]{\sf .L}=(\mathcal{L}_i)_1^{j-1}$\;
${\sf S}[i+L]{\sf .U}=1$\;
}
{\tt sortPDecrement}$({\sf S})$\;
\For{$i=1:L$}{
$(\mathcal{L}_i)_1^j=${\tt conc}$({\sf S}[i]{\rm .L},{\sf S}[i]{\rm .U})$\;
}
$\alpha=L$\;
}
}
$\hat{\mathbf{u}}=\mathbf{L}_1$\;
\end{algorithm}

Instead of making a hard decision for each information bit of $\mathbf{u}$ in the SC
algorithm, the SCL
algorithm creates two paths in which the information bit is assumed to be 0 and
1, respectively. If the number of paths is greater than the list size $L$, the
$L$ most reliable
paths are selected. At the end of the decoding procedure, the most reliable
path is chosen as $\hat{\mathbf{u}}$. The SCL algorithm
is formally described in Alg. \ref{alg:SCL}. Without loss of generality, we assume $L$ to
be a power of two, i.e.\ $L=2^l$. We use
$\mathbf{L}_i=((\mathcal{L}_i)_1,(\mathcal{L}_i)_1,\cdots,(\mathcal{L}_i)_{N})$
to represent the $i$-th list vector, where $0 < i \leq L$. ${\sf S}$ is a
structure type array with size $2L$. Each element of 
${\sf S}$ has three members: ${\sf P}$, ${\sf L}$, and ${\sf U}$. The function {\tt
  sortPDecrement} sorts the array ${\sf S}$ by decreasing order of ${\sf P}$. {\tt
  c=conc(a,b)} attaches a bit sequence {\tt b} at the end of a bit sequence {\tt
  a}, and the length of the output bit sequence 
{\tt c} is the sum of lengths of {\tt a} and {\tt b}.

The CA-SCL algorithm is used for the CRC-concatenated polar codes. The
difference between CA-SCL \cite{Tal2012} and SCL algorithms is how to
make the final decision for $\hat{\mathbf{u}}$. If there is at least one path
satisfying the CRC constraint, the most reliable CRC-valid path is chosen for
$\hat{\mathbf{u}}$. Otherwise, the decision rule of the SCL algorithm
is used for the CA-SCL algorithm. 
%Since now, without being specified, polar codes
%mentioned in the following sections are CRC-concatenated polar codes.

\section{$M$-bit Symbol-Decision Decoding Algorithms for Polar Codes}
\label{sec:SBDecoder}
\subsection{$M$-bit Symbol-Decision SC Algorithm}
Here, we proposed a symbol-decision SC algorithm, which treats $M$-bit data as a
symbol and decodes a symbol at a time. Let $\mathcal{Z}$ represent the alphabet
of all $M$-bit symbols. The symbol-decision SC algorithm deals with the virtual channel $\mathcal{W}_{N}^{(j)}:
\mathcal{Z}\rightarrow\mathcal{Y}^N\times\mathcal{Z}^{(j-1)}, 0 < j \leq
\frac{N}{M}$ with the transition probabilities $\mathcal{W}_{N}^{(j)}(y_1^N,z_1^{j-1}|z_j)$, where
$(y_1^N,z_1^{(j-1)})$ and $z_j=(u_{jM-M+1},\cdots,u_{jM})$ denote the output and input of
$\mathcal{W}_{N}^{(j)}$, respectively. Actually, $\mathcal{W}_{N}^{(j)}$ is
exactly equivalent to $W_{N,M}^{(j)}$ if we consider $\mathcal{X}^M$ as
the binary vector representation of $\mathcal{Z}$.
%For a polar code, given $y_1^N \text{ and
%} u_1^N$, $\mathcal{W}_{N}^{(j)}(y_1^N,z_1^{j-1}|z_j)$ is exactly equivalent to
%$W_{N,M}^{(j)}(\mathbf{y},u_1^{jM-M}|u_{jM-M+1}^{jM})$. 
Therefore, the symbol-decision SC algorithm has the same schedule as
the parallel SC algorithm in \cite{ParSC, ParSC2, ParSC3}. However, our symbol-decision SC algorithm has a different
approach, called symbol-based recursive channel combination, to compute
symbol-based channel transition probabilities $W_{N,M}^{(j)}(\mathbf{y},\hat{u}_1^{jM-M}|u_{jM-M+1}^{jM})$, which is our
main focus.

\subsection{Symbol-Based Recursive Channel Combination}
%An $M$-bit symbol-decision polar decoder contains $M$ component decoders. Each
%of them deals with a sub polar code with a length $\frac{N}{M}$.
Assume
$u_{iM-M+1}^{iM}=(w_i,w_{i+\frac{N}{M}},\cdots,w_{i+N-\frac{N}{M}})B_MF^{\otimes
  m}$ for $1 \leq i \leq \frac{N}{M}$. In \cite{ParSC, ParSC2, ParSC3}, the calculation of the symbol-based channel
transition probability $W_{N,M}^{(i)}(\mathbf{y},\hat{u}_1^{iM-M}|u_{iM-M+1}^{iM})$ is
based on the following equation, referred to as direct-mapping calculation:
\begin{equation}
\label{eq:DCS}
\begin{split}
W&_{N,M}^{(i)}(\mathbf{y},\hat{u}_1^{iM-M}|u_{iM-M+1}^{iM}) = \\
&\prod_{j=0}^{M-1}W_{\frac{N}{M},1}^{(i)}(y_{j\frac{N}{M}+1}^{(j+1)\frac{N}{M}},\hat{w}_{1+j\frac{N}{M}}^{(i-1)+j\frac{N}{M}}|w_{i+j\frac{N}{M}}),
\end{split}
\end{equation}
where
$W_{\frac{N}{M},1}^{(i)}(y_{j\frac{N}{M}+1}^{(j+1)\frac{N}{M}},\hat{w}_{1+j\frac{N}{M}}^{(i-1)+j\frac{N}{M}}|w_{i+j\frac{N}{M}})$
is calculated by the Arikan's recursive channel transformations.

Actually, a symbol-based
recursive channel combination described in Proposition~\ref{prop:1} can be used
to calculate $W_{N,M}^{(i)}(\mathbf{y},\hat{u}_1^{iM-M}|u_{iM-M+1}^{iM})$.

%If all bits of $\mathbf{u}$ are independent and each bit has an equal probability
%of being a 0 or 1, the following symbol-based recursive channel combination
%relationship can be used to calculate the symbol-based channel transition
%probability $\Pr(\mathbf{y},u_0^{jM-1}|u_{jM}^{jM+M-1})$:  
\begin{PROP}
\label{prop:1}
Assume that all bits of $\mathbf{u}$ are independent and each bit has an equal probability
of being a 0 or 1. Given $0 < m \leq n$, $N=2^n$, $M=2^m$, for any $1 \leq \phi
\leq m$, $0 \leq \lambda <n$, $\Lambda=2^{\lambda}$, $\Phi=2^{\phi}$, and $0
\leq i < \frac{2\Lambda}{\Phi}$, we say that a $\Phi$-bit channel
$W_{2\Lambda,\Phi}^{(i+1)}$ is obtained by a single-step combination of two
independent copies of a $\frac{\Phi}{2}$-bit channel $W_{\Lambda,\Phi/2}^{(i+1)}$ and write
\begin{equation}
(W_{\Lambda,\Phi/2}^{(i+1)},W_{\Lambda,\Phi/2}^{(i+1)}) \mapsto W_{2\Lambda,\Phi}^{(i+1)},
\end{equation}
where the channel transition probability satisfies,
\begin{equation}
\label{eq:prop1}
\begin{split}
W&_{2\Lambda,\Phi}^{(i+1)}(y_{1}^{2\Lambda},u_1^{i\Phi}|u_{i\Phi+1}^{i\Phi+\Phi}) = \\
&W_{\Lambda,\Phi/2}^{(i+1)}(y_1^{\Lambda},u_{1,o}^{i\Phi}\oplus u_{1,e}^{i\Phi}|u_{i\Phi+1,o}^{i\Phi+\Phi}\oplus u_{i\Phi+1,e}^{i\Phi+\Phi})\\
&\cdot W_{\Lambda,\Phi/2}^{(i+1)}(y_{\Lambda+1}^{2\Lambda},u_{1,e}^{i\Phi}|u_{i\Phi+1,e}^{i\Phi+\Phi}).
\end{split}
\end{equation}
\end{PROP}

%We call Eq.~\eqref{eq:prop1} as the symbol-based recursive channel
%combination relationship. 

Similar to the SC algorithm, with the help of the symbol-based recursive channel
combination, an $M$-bit symbol-decision SC
algorithm can be represented by using a message flow graph (MFG) as well, where
a channel transition probability is referred to as a {\it message} for the sake of
convenience. This MFG is referred to as SR-MFG. If the code length of a polar
code is $N$, the SR-MFG can be divided into $(n+1)$ stages ($\text{S}_0, \text{S}_1, 
\cdots, \text{S}_n$) from the right to the left: one initial stage $\text{S}_0$ and $n$ calculation stages. For the SC
algorithm, all calculation stages carry out the Arikan's recursive channel transformation. However, for the $M$-bit symbol-decision SC
algorithm, in the left-most $m$ calculation
stages ($\text{S}_n, \cdots, \text{S}_{n-m+1}$), called S-COMBS stages, 
symbol-based channel combinations are carried out. For the rest $(n-m)$
calculation stages ($\text{S}_{n-m}, \cdots, \text{S}_1$), called B-TRANS
stages, the Arikan's recursive channel transformations are performed. The
S-COMBS stages use outputs of B-TRANS stages to calculate symbol-based 
messages. 

For \cite{ParSC, ParSC2, ParSC3}, we refer to the MFG as the DM-MFG which also
consists of two parts: B-TRANS and DM-CAL. The B-TRANS 
part of the DM-MFG is the same as that of the SR-MFG. However, there is only
one stage in the DM-CAL part of the DM-MFG which performs the direct-mapping
calculation. 

\begin{figure}[htbp]
\centering
\includegraphics[width=6cm]{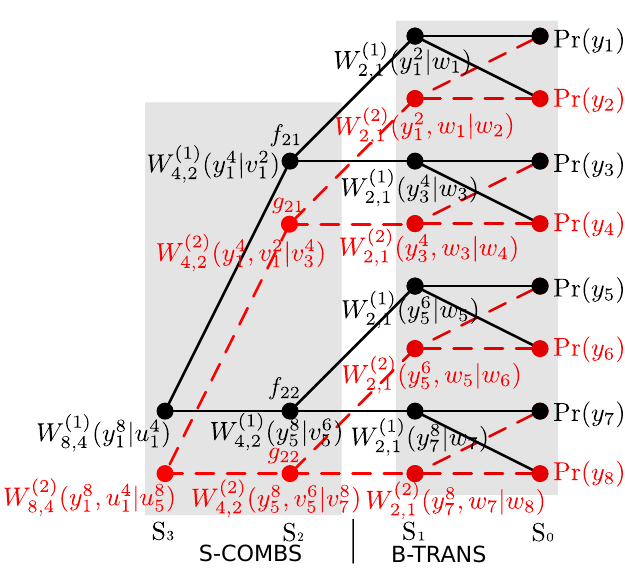}
\caption{The message flow graph of a four-bit symbol-decision SC algorithm
  for a polar code with a code length of eight by using the proposed symbol-based
  recursive channel combination.}
\label{fig:MFG_8_4}
\end{figure}

\begin{figure}[htbp]
\centering
\includegraphics[width=5.8cm]{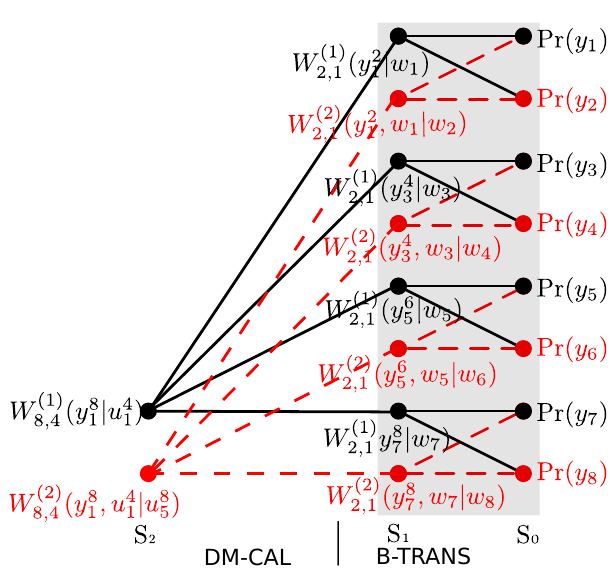}
\caption{The message flow graph of a four-bit symbol-decision SC algorithm
  for a polar code with a code length of eight by using direct-mapping
  calculation \cite{ParSC, ParSC2, ParSC3}.}
\label{fig:MFG_8_4_DM}
\end{figure}

For example, as shown in Fig.~\ref{fig:MFG_8_4}, the SR-MFG of a
four-bit symbol-decision SC algorithm for a polar code with $N=8$  has four stages. Messages of the initial stage ($\text{S}_0$) come
from the channel directly. Messages of the first stage ($\text{S}_1$) are
calculated with Arikan's transformations. Messages of the second
and third stages (S2 and S3) are calculated with Eq.~\eqref{eq:prop1}. Stages in the
left gray box are the S-COMBS stages. Stages in the right gray box are the B-TRANS
stages. Fig.~\ref{fig:MFG_8_4_DM} shows the DM-MFG when the direct-mapping
calculation is used to calculate symbol-based channel transition probability
$W_{8,4}^{(1)}(y_1^8|u_1^4)$ and $W_{8,4}^{(2)}(y_1^8,u_1^4|u_5^8)$. Here,  
\begin{align*}
v_1^4&=u_{1,o}^8\oplus u_{1,e}^8, \hspace{4mm} v_5^8=u_{1,e}^8,\\
w_1&=v_1\oplus v_2=u_1\oplus u_2 \oplus u_3 \oplus u_4,\\
w_2&=v_3\oplus v_4=u_5\oplus u_6 \oplus u_7 \oplus u_8,\\
w_3&=v_2=u_3\oplus u_4,\\
w_4&=v_4=u_7\oplus u_8,\\
w_5&=v_5\oplus v_6 = u_2\oplus u_4,\\
w_6&=v_7\oplus v_8 = u_6 \oplus u_8,\\
w_7&=v_6 = u_4,\\
w_8&=v_8 = u_8. 
\end{align*}

For the direct-mapping calculation, Eq.~\eqref{eq:DCS} needs $(M-1)$ additions. Therefore, a total of
$2^{\lvert\mathcal{AM}_j\rvert}(M-1)$ additions are needed to calculate all LL-based symbol-based channel
transition probabilities for $u_{jM+1}^{jM+M}$. Consider the recursive
symbol-based channel combination. The S-COMBS stages of the SR-MFG are indexed as
1 to $m$ from left to right. There are $2^{n-i} (0 < i \leq m)$ nodes in the $i$-th
S-COMBS stage and each node contains $2^{M+i-n}$ messages. One addition is
needed to compute each LL message according to 
Eq.~\eqref{eq:prop1}. Hence, the number of additions needed by the S-COMBS stages to
calculate $W_{N,M}^{(j)}(\mathbf{y},\hat{u}_1^{jM-M}|u_{jM-M+1}^{jM})$ is
$\sum_{i=1}^{m-1}2^{i}2^{\frac{M}{2^{i}}}+2^{\lvert\mathcal{AM}_j\rvert}$. Actually,
if we perform the hardware implementation, the worst case - that all bits of a
symbol are information bits - should be considered. Therefore, the recursive 
symbol-based channel combination can be taken advantage of to reduce
complexity of calculating the symbol-based channel transition probability.  

For the example shown in Fig.~\ref{fig:MFG_8_4}, Eq.~\eqref{eq:DCS} needs $2^4(4-1)=48$
additions to calculate $\log(W_{8,4}^{(1)}(y_1^8|u_1^4))$. With the symbol-based channel
combination, $4$, $4$ and $16$ additions are needed to calculate
$\log(W_{4,2}^{(1)}(y_1^4|v_1^2))$, $\log(W_{4,2}^{(1)}(v_5^8|v_5^6))$ and $\log(W_{8,4}^{(1)}(y_1^8|u_1^4))$,
respectively. Therefore, our method needs only $2^4+2\times 2^2=24$
additions, which is only a half of those needed by
Eq.~\eqref{eq:DCS}. Table~\ref{tab:adder_comp} lists the numbers of additions
needed by our recursive method and direct-mapping calculation \cite{ParSC,
  ParSC2, ParSC3} when all $M$ bits of a symbol are information bits. When $M=8$, the number of additions needed by our proposed method is 17\% of that
needed by the direct-mapping calculation.

\begin{table}[hbtp]
\begin{center}
\caption{The numbers of additions to calculate
  $W_{N,M}^{(j+1)}(\mathbf{y},\hat{u}_1^{jM}|u_{jM+1}^{jM+M})$ when the $(j+1)$-th
symbol has no frozen bit.}
\label{tab:adder_comp}
\begin{tabular}{|c|c|c|}
\hline
 & Proposed method & Direct-mapping calculation \cite{ParSC, ParSC2, ParSC3} \\ \hline
%$M=2^m$ & $\sum_{i=0}^{m-1}2^{i}2^{\frac{M}{2^{i}}}$ & $2^M(M-1)$  \\ \hline\hline
$M=2$ & 4 & 4  \\ \hline
$M=4$ & 24 & 48  \\ \hline
$M=8$ & 304 & 1792  \\ \hline
\end{tabular}
\end{center}
\end{table} 

The other advantage of the proposed method to calculate the symbol-based channel
transition probability is that it reveals the similarity between the Arikan's
recursive channel transformation and symbol-based
recursive channel combination. We will take advantage of this similarity to
reuse adders and to save area when computing the bit- and symbol-based channel transition
probability in our proposed architecture. In \cite{ParSC2}, additional dedicated adders are used to
calculated the symbol-based channel transition probability, which is not area efficient.

%In terms of the error performance, simulations of \cite{ParSC, ParSC2} show that there
%is no observed performance loss for the $M$-bit symbol-decision SC/SCL
%algorithm using Eq.~\eqref{eq:DCS} to calculate the symbol-based message, compared with the SC/SCL
%algorithm. Since our proposed method performs the same function
%as Eq.~\eqref{eq:DCS} used in \cite{ParSC}, the $M$-bit symbol-decision SC/SCL algorithm using our
%symbol-based recursive channel transformation should have the same performance as the $M$-bit
%symbol-decision SC/SCL algorithm in \cite{ParSC, ParSC2}. Consequently, our proposed
%method should not introduce any observed performance degradation to the $M$-bit
%symbol-decision SC/SCL algorithm compared with the SC/SCL algorithm. This is
%consistent with our simulation results. 

\begin{figure}[htbp]
\centering
\includegraphics[width=7cm]{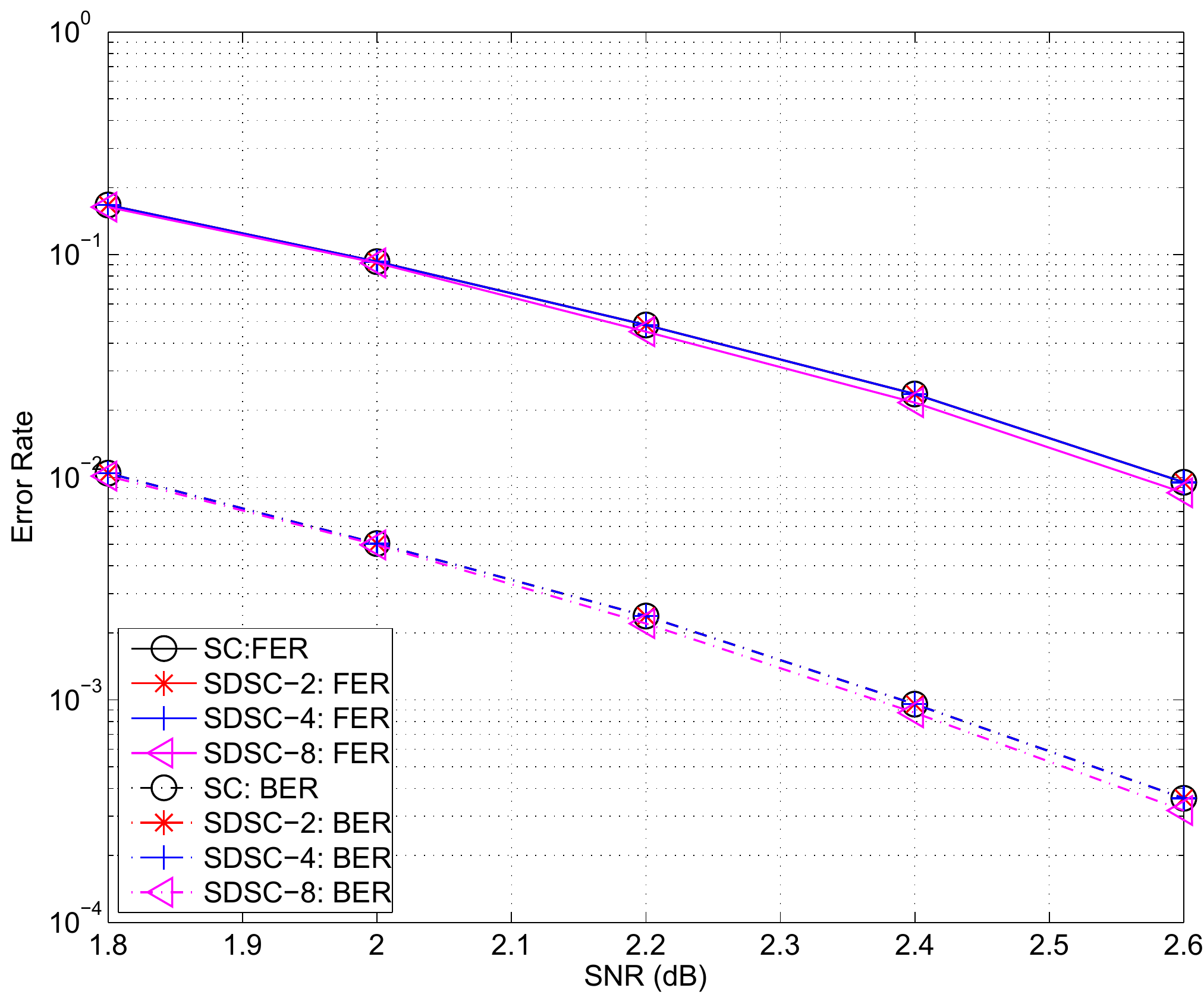}
\caption{Error rates of symbol-decision SC algorithms for a
  (1024, 512) polar code.}
\label{fig:SDSC_PER}
\end{figure}

In terms of the error performance, the symbol-decision SC algorithm is not
worse than the bit-decision SC algorithm \cite{FER_performance}. Fig.~\ref{fig:SDSC_PER} shows the BERs
and FERs of symbol-decision SC algorithms for a (1024, 512) polar
codes. SDSC-$i$ denotes the $i$-bit symbol-decision SC algorithm. When $M=2$ and
$4$, the FER performance is the same as that of the bit-decision
SC algorithm. When $M=8$, the FER performance is slightly better. 
%Therefore, by applying the symbol-based recursive channel
%transformation, the symbol-decision SC/SCL algorithms have no observed performance
%loss compared with the bit-decision SC/SCL algorithm. Even with different symbol
%size $M$, these performance curves are very close. 

\subsection{Generalized Symbol-Decision SCL Decoding Algorithm}
%The symbol-based recursive channel combination is also useful for SCL algorithms to calculate the path
%metrics. In the bit-decision SCL algorithm, for each information bit,
%the path expansion coefficient is two. Our symbol-decision SCL algorithm is more complicated than the
%bit-decision SCL algorithm, since the path expansion coefficient is not a
%constant any more. Therefore, we also focus on path pruning for the
%symbol-decision SCL algorithm.

Similarly, the symbol-based recursive channel combination is also useful for the
SCL algorithm. The symbol-decision SCL algorithm is more complicate than the SCL algorithm, since
the path expansion coefficient is not a constant any more. In the SCL algorithm, for each
information bit, the path expansion coefficient is two. But for the $M$-bit
symbol-decision SCL algorithm, the path expansion coefficient is
$2^{\lvert\mathcal{AM}_j\rvert}$, which depends on
the number of information bits in an $M$-bit symbol. The $M$-bit symbol-decision
SCL algorithm is formally described in Alg.~\ref{alg:SBSCL}.
Without any ambiguity, $\mathbf{0}$ represents a zero vector whose bit-width
is determined by the left-hand operator. The function {\tt dec2bin}$(d,b)$
converts a decimal number $d$ to a $b$-bit binary vector. Eq.~\eqref{eq:prop1}
is used to calculate the symbol-based channel transition
probability corresponding to each list, i.e.\ $W_{N,M}^{(j+1)}(\mathbf{y},(\mathcal{L}_i)_1^{jM}|u_{jM+1}^{jM+M})$.

\begin{algorithm}
\caption{$M$-bit Symbol-Decision SCL Decoding Algorithm}
\label{alg:SBSCL}
\LinesNumbered
$\alpha=1$\;
\For{$j=1:\frac{N}{M}$}{
$\beta=2^{\lvert\mathcal{AM}_j\rvert}$\;
\uIf{$\beta==1$}{
\For{$i=1:\alpha$}{
$(\mathcal{L}_i)_{jM-M+1}^{jM} = \mathbf{0}$\;
}
}
\uElseIf{$\alpha\beta\leq L$}{
$u_{\mathcal{AM}_j^c}=\mathbf{0}$\;
\For{$k=0:\beta-1$}{
$u_{\mathcal{AM}_j}=${\tt dec2bin}$(k,\lvert\mathcal{AM}_j\rvert)$\;
\For{$i=1:\alpha$}{
$t=i+k\alpha$\;
$(\mathcal{L}_{t})_{1}^{jM} = ${\tt conc}$((\mathcal{L}_i)_{1}^{jM-M},u_{jM-M+1}^{jM})$\;
}
}
$\alpha=\alpha\beta$\;
}
\Else{
$u_{\mathcal{AM}_j^c}=\mathbf{0}$\;
\For{$k=0:\beta-1$}{
$u_{\mathcal{AM}_j}=${\tt dec2bin}$(k,\lvert\mathcal{AM}_j\rvert)$\;
\For{$i=1:L$}{
$t=i+kL$\;
${\sf S}[t]{\sf .P}=W_{N,M}^{(j)}(\mathbf{y},(\mathcal{L}_i)_1^{jM-M}|u_{jM-M+1}^{jM})$\;
${\sf S}[t]{\sf .L}=(\mathcal{L}_i)_1^{jM-M}$\;
${\sf S}[t]{\sf .U}=u_{jM-M+1}^{jM}$\;
}
}
{\tt sortPDecrement}({\sf S})\;
\For{$i=1:L$}{
$(\mathcal{L}_i)_1^{jM}=${\tt conc}$({\sf S}[i]{\rm .L},{\sf S}[i]{\rm .U})$\;
}
$\alpha=L$\;
}
}
\end{algorithm}

\begin{figure}[htbp]
\centering
\includegraphics[width=7cm]{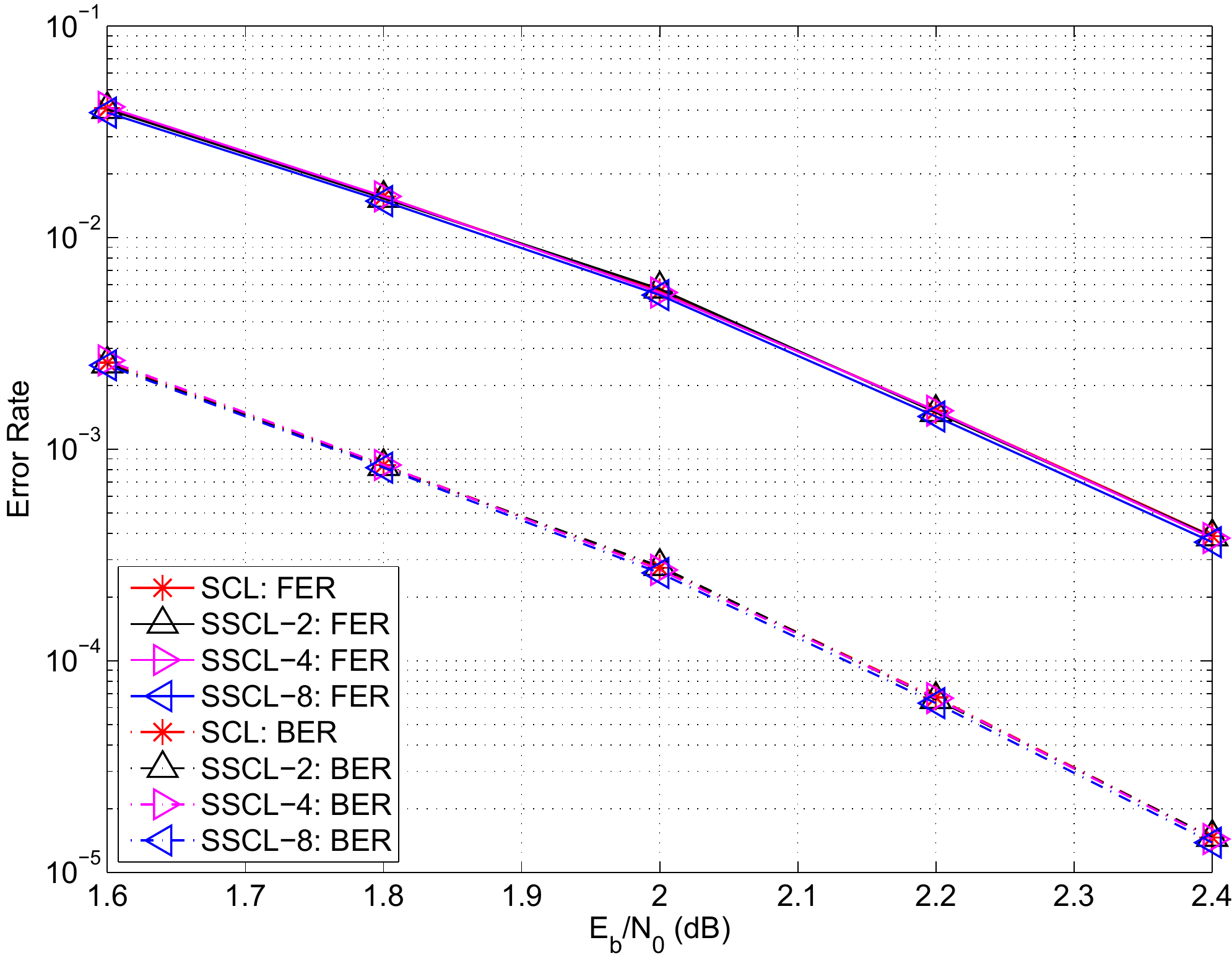}
\caption{Error rates of symbol-decision SCL algorithms for a
  (1024, 480) CRC32-concatenated polar code with $L=4$.}
\label{fig:SBSCL_PER}
\end{figure}

Fig.~\ref{fig:SBSCL_PER} shows the BERs and FERs of symbol-decision SCL
algorithms for a (1024, 480) CRC32-concatenated polar code with $L=4$ where the
generator polynomial of the CRC32 is 0x1EDC6F41. This CRC32 is also used in all
the CRC-concatenated polar codes used in the following section. SDSCL-$i$
denotes the $i$-bit symbol-decision SCL algorithm. The performances
of the symbol-decision SCL algorithms with different symbol
sizes are almost the same.

\subsection{Two-Stage List Pruning Network for Symbol-Decision SCL algorithm}
\label{sec:TSLPN}
\begin{figure}[htbp]
\centering
\includegraphics[width=7.5cm]{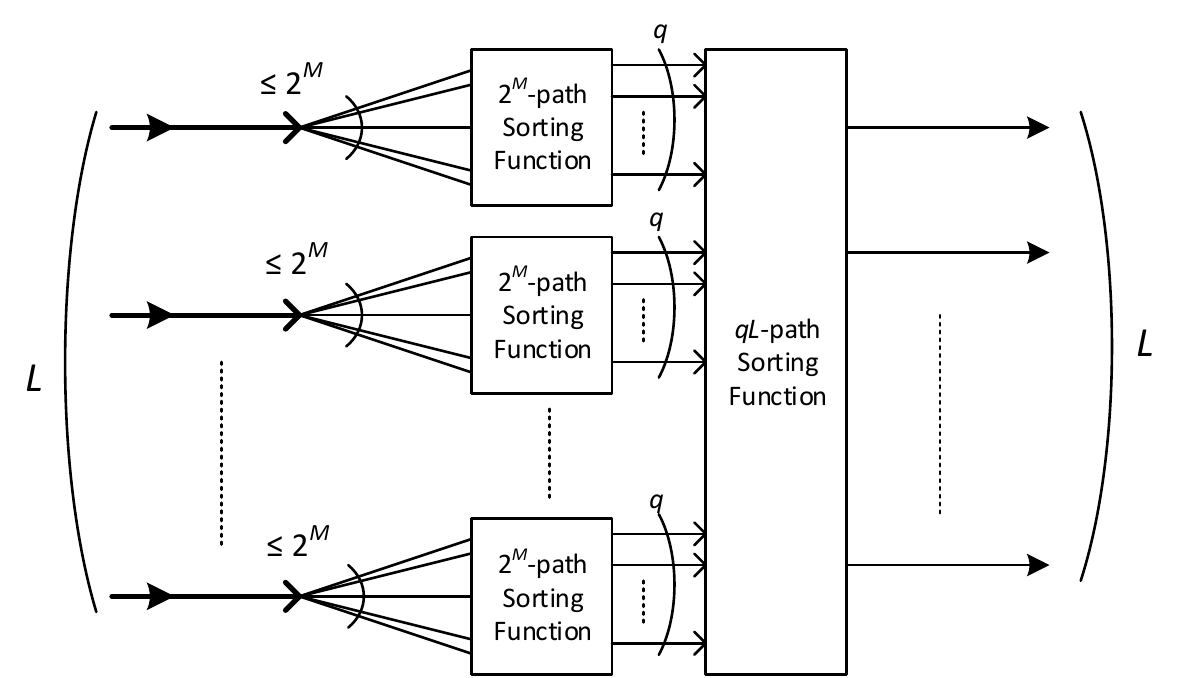}
\caption{Message flow for a two-stage list pruning network.}
\label{fig:TSLPP}
\end{figure}

For the $M$-bit
symbol-decision SCL algorithm, the maximum path expansion
coefficient is $2^M$, i.e.\, each existing path generates $2^M$ paths. Therefore, in the worst-case scenario, the $L$ most reliable paths should be selected 
out of $2^ML$ paths. To facilitate this sorting network, we propose
a two-stage list pruning network. In the first stage, the $q$ most reliable paths
are selected from up to $2^M$ paths that come from expansion of each existing path. Therefore, there are $qL$ paths left. In the second stage, the $L$ most reliable paths are sorted out
from the $qL$ paths generate by the first stage. The message flow of a two-stage list pruning
network is illustrated in Fig.~\ref{fig:TSLPP}. 
 
If $q \geq L$, the $L$ paths found by the two-stage list
pruning network are exactly the $L$ most reliable paths among the $2^ML$
paths. When $q < L$, the probability that the $L$ paths found by the two-stage list
pruning network are exactly the $L$ most reliable paths among the $2^ML$ paths
decreases as well. This may cause some performance loss. But a smaller $q$ leads to a two-stage list pruning
network with lower complexity.

\begin{figure}[htbp]
\centering
\includegraphics[width=7.5cm]{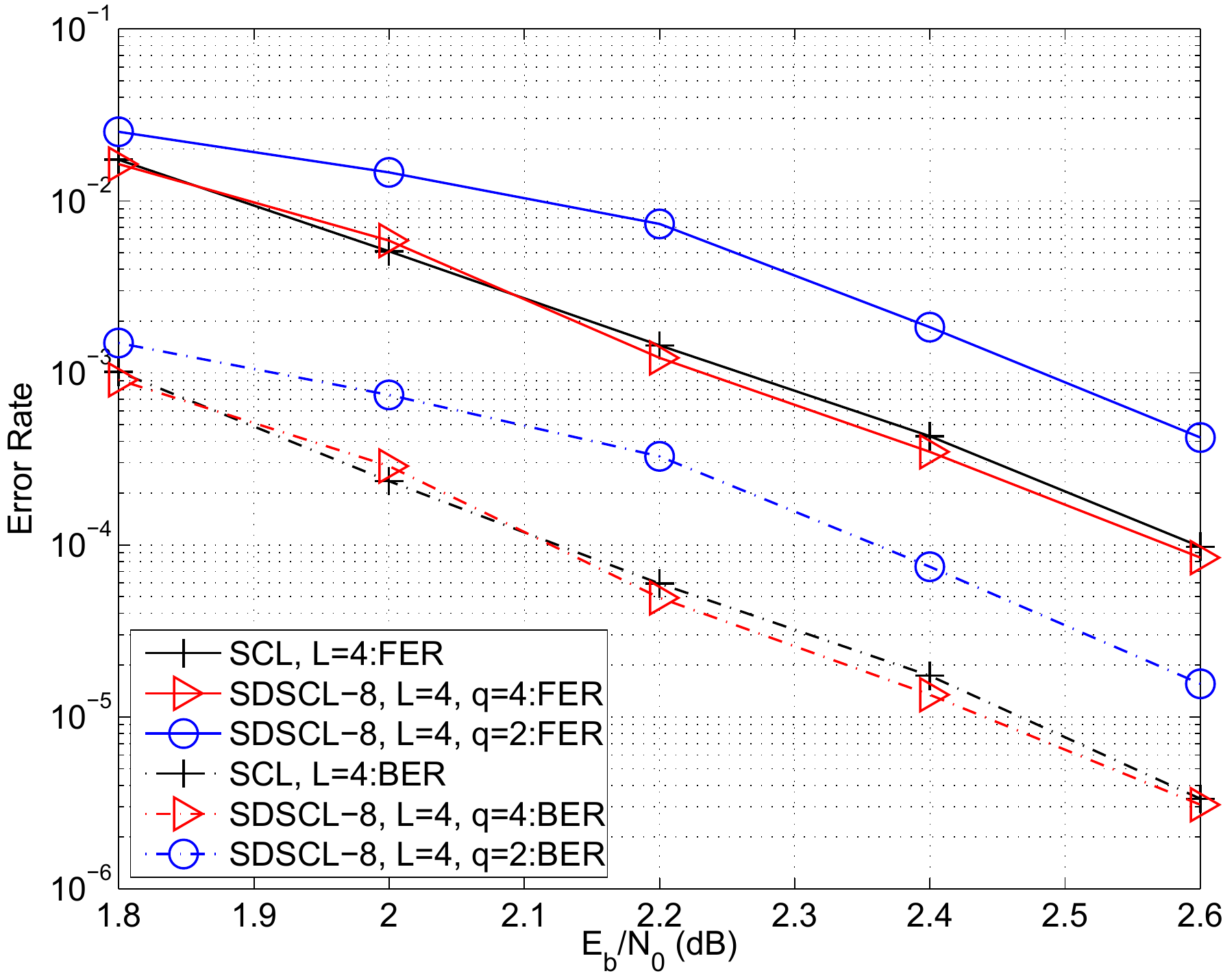}
\caption{Error rates of the SDSCL-8 decoder for a
  (1024, 480) CRC32-concatenated polar code with $L=4$.}
\label{fig:TSLPP_4}
\end{figure}

\begin{figure}[htbp]
\centering
\includegraphics[width=7.5cm]{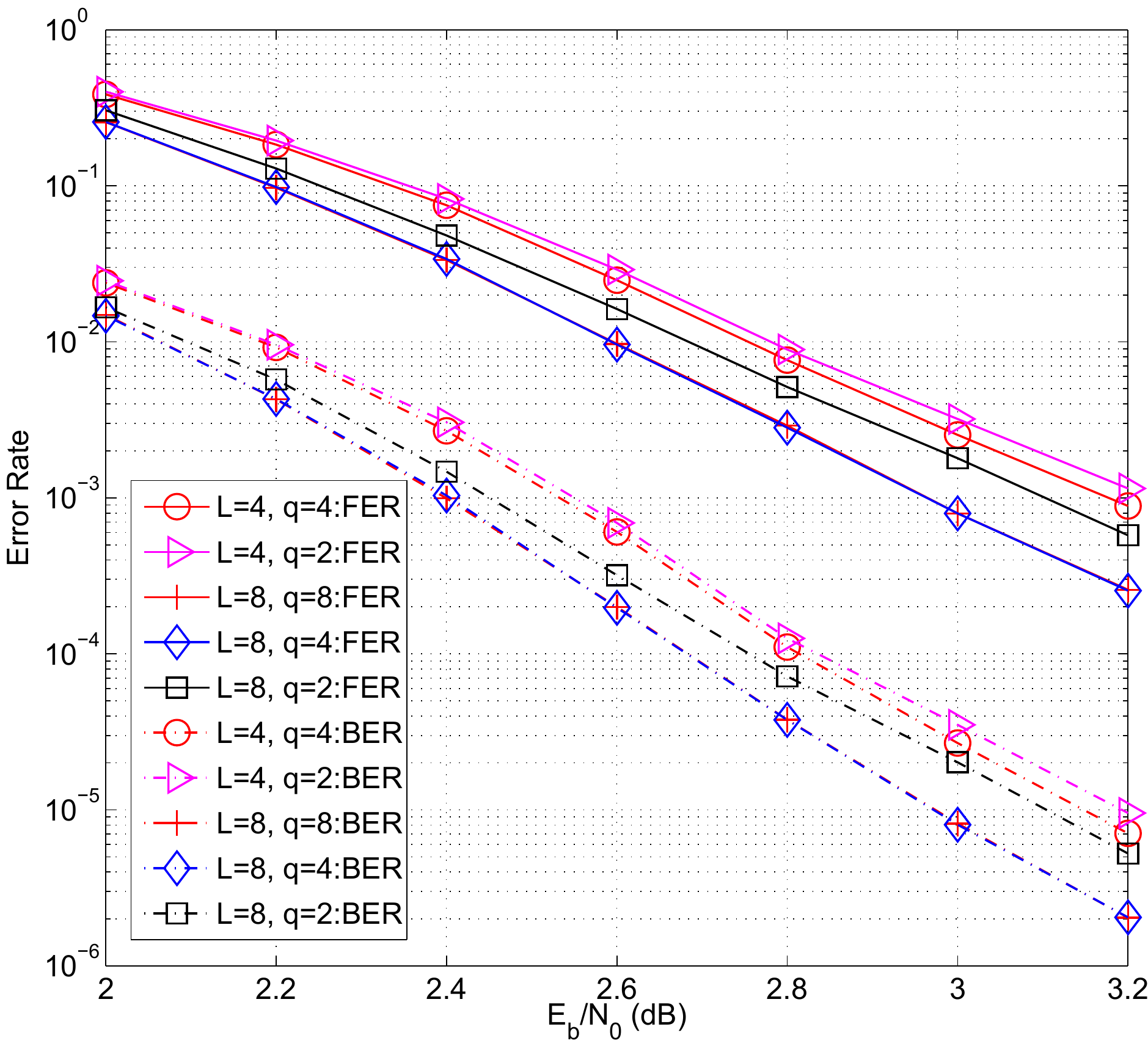}
\caption{Error rates of the SDSCL-4 algorithm for a
  (2048, 1401) CRC32-concatenated polar code with $L=4$ and $L=8$.}
\label{fig:TSLPP_8_2048}
\end{figure}

Fig.~\ref{fig:TSLPP_4} shows how different values of $q$ affect the error performance of an
SDSCL-8 algorithm for a (1024, 480) CRC32-concatenated polar code with $L=4$. When $L=4$ and $q=2$, the SDSCL-8 algorithm
shows an FER performance loss of about 0.25 dB at an FER level of $10^{-3}$.
As shown in Fig.~\ref{fig:TSLPP_8_2048}, for a (2048,1401)
CRC32-concatenated polar
code, the two stage list-pruning network of $q=4$ helps to reduce 
the complexity of the SDSCL-4 decoder without observed performance loss when
$L=8$. When $q=2$ and $L=8$, the SDSCL-4 decoder has a performance degradation of
about 0.1 dB at an FER level of $10^{-3}$, compared with the SDSCL-4 decoder with
$q=8$ and $L=8$. If $L=4$, the error performance due to $q=2$ is very small.

Therefore, the two-stage list pruning network uses an additional parameter $q$ to
introduce different trade-offs between error performance and complexity.

\section{Pre-Computation Memory-Saving Technique}
\label{sec:PCMS}
Pre-computation technique was first proposed in \cite{135746} and can be used
to improve processing rate when the number of possible outputs is finite. In
\cite{6475198}, the pre-computation technique is used to improve the throughput
of the line SC decoder with an additional cost of increased area. Here, our main
purpose is to use the pre-computation technique to reduce the memory required by
list decoders because the memory of an SCL decoder to store the channel
transition probability becomes a big challenge as the list size and code length
increase. Henceforth, this memory saving technique is called the pre-computation
memory-saving (PCMS) technique. It is worth noting that this
memory-saving technique is \emph{independent} of the decoder architecture and the
message representation of SCL decoders.

Let us take the MFG shown in Fig.~\ref{fig:MFG_8_4} as an example. For stages $\text{S}_0$ and $\text{S}_1$, the numbers of pairs of LLs stored by the list decoder are 8
and $4L$, respectively. Actually, the outgoing message $W_{2,1}^{(1)}(y_1^2|w_1)$ of the top
black node in $\text{S}_1$ can only be either $W_{2,1}^{(1)}(y_1^2|0)$ or $W_{2,1}^{(1)}(y_1^2|1)$. The
outgoing message $W_{2,1}^{(2)}(y_1^2,w_1|w_2)$ can only be one of $W_{2,1}^{(2)}(y_1^2,0|0)$,
$W_{2,1}^{(2)}(y_1^2,0|1)$, $W_{2,1}^{(2)}(y_1^2,1|0)$, and
$W_{2,1}^{(2)}(y_1^2,1|1)$. Hence, no matter what the list size is, the total number of
possible values of outgoing messages of $\text{S}_1$ is $2\times 4+4\times 4=24$. These 24
values provide all information we need for calculations of further stages. With
knowledge of these 24 values, channel LLs are not needed any more.

Generally speaking, the PCMS technique takes advantage of the relationship between
messages of $\text{S}_0$ (channel LLs), and outgoing messages of
$\text{S}_1$. By storing only all possible outgoing messages of $\text{S}_1$, the PCMS technique helps list
decoders save memory. 

Let us evaluate the memory saving of the PCMS technique, assuming LL
representation is used for the channel transition probability. Without PCMS
technique, a list decoder for a polar code with the code length of $N$ has a
list size of $L$ stores $(N-2)L+N$ LL pairs. Each pair contains
two messages which are associated with the conditional bit being zero or
one. The total number of bits used for LL storage is
\begin{equation}
\begin{split}
B_{\mbox{LL}}&=2\Bigl(NQ_{ch}+L\sum_{i=1}^{\log N-1}2^i(Q_{ch}+\log N-i)\Bigr) \\
&=2(L+1)NQ_{ch}+4L(N-\log N-Q_{ch}-1),
\end{split}
\end{equation}
where $Q_{ch}$ denotes the number of bits used for the quantization of the channel LLs. 

With the PCMS technique, the total number of LL pairs needed
by a list decoder is $\frac{N}{2}L+\frac{3}{2}N$. The total number of bits
needed for LL storage is:

\begin{equation}
\begin{split}
B_{\mbox{PCMS}}=&2(\frac{N}{2}+N)(Q_{ch}+1)+\\
&2L\sum_{i=1}^{\log N-2}2^i(Q_{ch}+\log N-i)\\
=&3N(Q_{ch}+1)+LN(Q_{ch}+3)\\&-4L(\log N+Q_{ch}+1)\\
=&B_{\mbox{LL}}-N(LQ_{ch}+L-Q_{ch}-3).
\end{split}
\end{equation}

Therefore, when LL representation is used for messages, the PCMS technique saves
$N(LQ_{ch}+L-Q_{ch}-3)$ bits of memory. The saving is linear with both $N$ and
$L$. Consider a polar code with $N=1024$, a list decoder with $L=4$ and
$Q_{ch}=4$. Without the PCMS technique, $B_{\mbox{LL}} = 57104$. With the PCMS
technique, $B_{\mbox{PCMS}}=43792$. The PCMS technique helps to save 13312 bits
of memory, which is 23\% of $B_{\mbox{LL}}$.  

The other advantage of the PCMS technique is that it improves the throughput
slightly because the messages of $\text{S}_1$ are already in the memory and
don't need to be calculated from the channel messages. For example, for a bit-decision
semi-parallel SCL decoder with the list size of $L$, if the code length is $N$
and the number of processing units is $P$, the latency saving due to the PCMS
technique is $\frac{NL}{P}$ clock cycles.

\section{Implementation of Symbol-Decision SCL Decoders}
\label{sec:ImpSBDecoder}
\subsection{Architecture of Symbol-Decision SCL Decoders}
\begin{figure}[htbp]
\centering
\includegraphics[width=9cm]{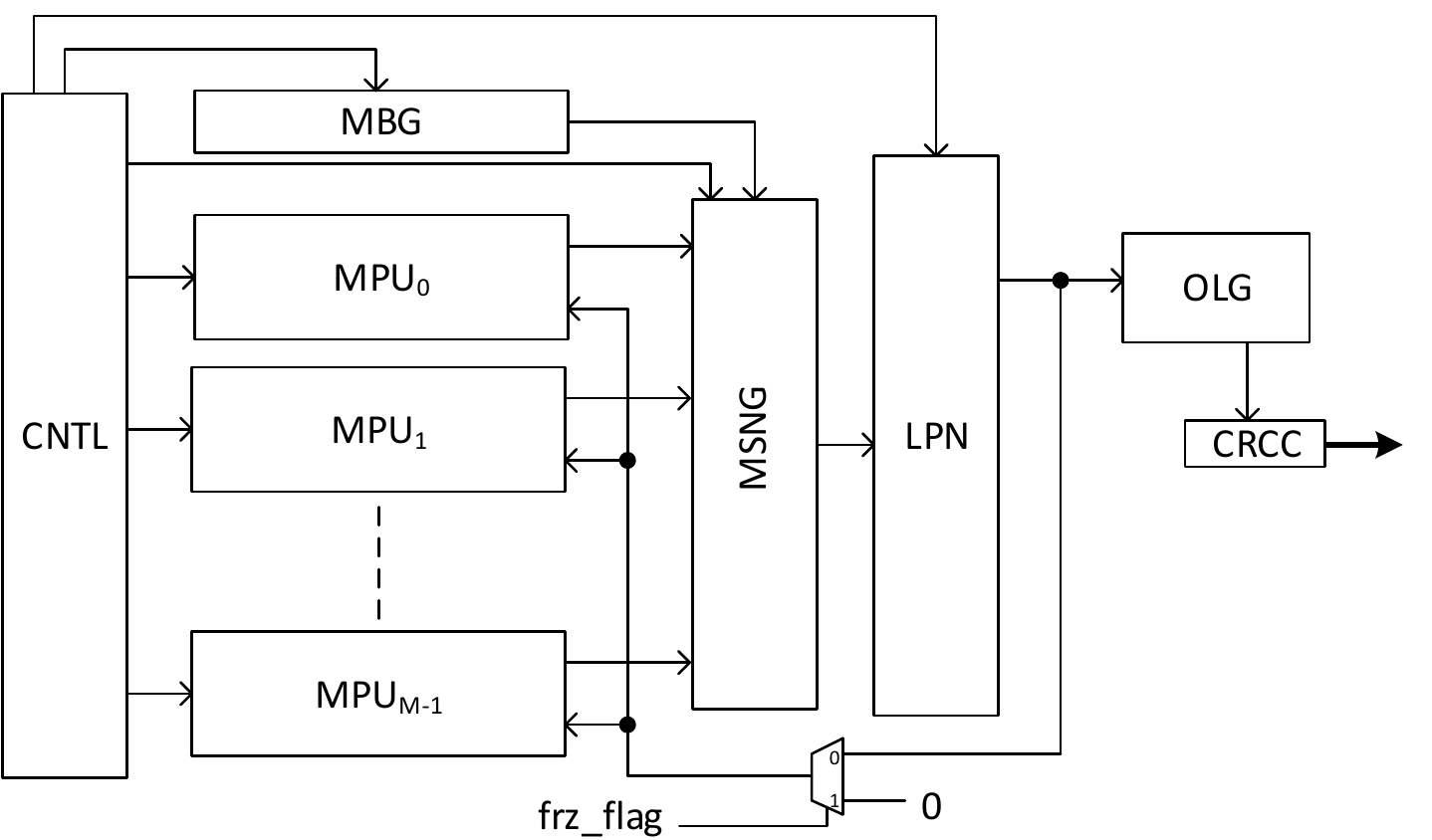}
\caption{Top architecture for an $M$-bit symbol-decision SCL decoder.}
\label{fig:SBSCL_arch}
\end{figure}

We propose an architecture of an $M$-bit symbol-decision SCL decoder shown in
Fig.~\ref{fig:SBSCL_arch}. It consists of $M$ MPU blocks ($\text{MPU}_0,
 \text{MPU}_1, \cdots , \text{MPU}_{M-1}$), a list
pruning network (LPN), a mask bit generator (MBG), a message-screening block
(MSNG), a control block (CNTL), an output-list generator (OLG) and a CRC checker (CRCC).

\begin{figure}[htbp]
\centering
\includegraphics[width=9cm]{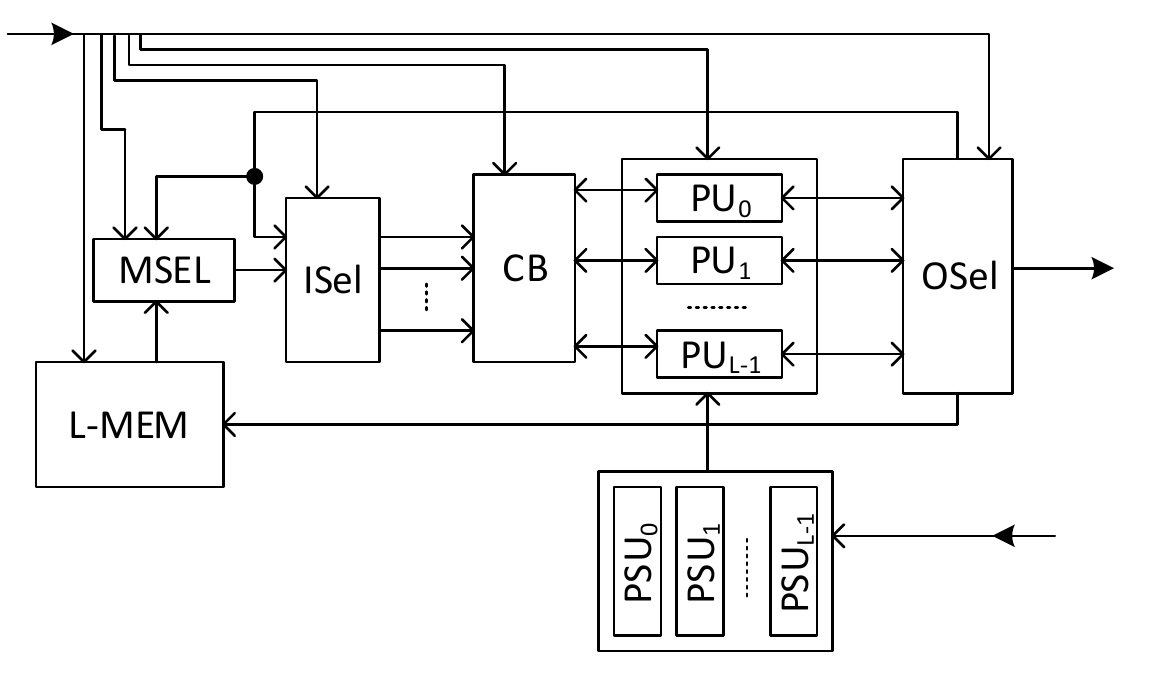}
\caption{Architecture of an MPU.}
\label{fig:SDSCL_arch}
\end{figure}

An MPU block calculates messages for B-TRANS and S-COMBS messages 
and updates the partial-sum network by adopting blocks of the SCL decoder in \cite{JunPolarList}. The additions of S-COMBS stages are carried
out by reusing the same hardware resource which is used to calculate messages of
B-TRANS stages to reduce the
area. Compared with the SCL decoder in \cite{JunPolarList}, the MPU has neither path pruning unit nor
the CRC checker. The other improvement for the MPU is that PCMS
technique is used here. The architecture of an MPU is shown in
Fig.~\ref{fig:SDSCL_arch}. Channel messages are not needed any more due to the adoption of PCMS
technique. L-MEM stores messages corresponding to stages of the MFG. For the stage
$\text{S}_1$, MSEL selects the appropriate messages from L-MEM based on partial sum values
and/or the type of calculation nodes. PUs are processing units to calculate LL
messages. PSUs is used to update partial-sums. ISel
selects messages from LMEM or OSel module for the crossbar (CB) module
which chooses proper messages for PUs. OSel outputs messages to L-MEM for
intermediate stages and output symbol-based messages to MSNG.

We take the MFG of Fig.~\ref{fig:MFG_8_4} as an example to illustrate the
function of block MSEL. For node $f_{21}$ of path $l$,
$\{W_{2,1}^{(1)}(y_1^2|0),W_{2,1}^{(1)}(y_1^2|1)\}_l$ and $\{W_{2,1}^{(1)}(y_3^4|0),W_{2,1}^{(1)}(y_3^4|1)\}_l$ are selected from LMEM by MSEL and
output to Isel. For node $g_{21}$ of path $l$,
$\{W_{2,1}^{(2)}(y_1^2,{w_1}_l|0),W_{2,1}^{(2)}(y_1^2,{w_1}_l|1)\}_l$ and
$\{W_{2,1}^{(2)}(y_3^4,{w_3}_l|0),W_{2,1}^{(2)}(y_3^4,{w_3}_l|1)\}_l$ are selected from
LMEM. Here, ${w_1}_l$ and ${w_3}_l$ are the partial sum for $w_1$ and $w_3$,
respectively, belonging to path $l$. The detailed information of other blocks in
Fig.~\ref{fig:SDSCL_arch} can be found in \cite{JunPolarList} and will not be
discussed in this paper.

\begin{figure}[htbp]
\centering
\includegraphics[width=8cm]{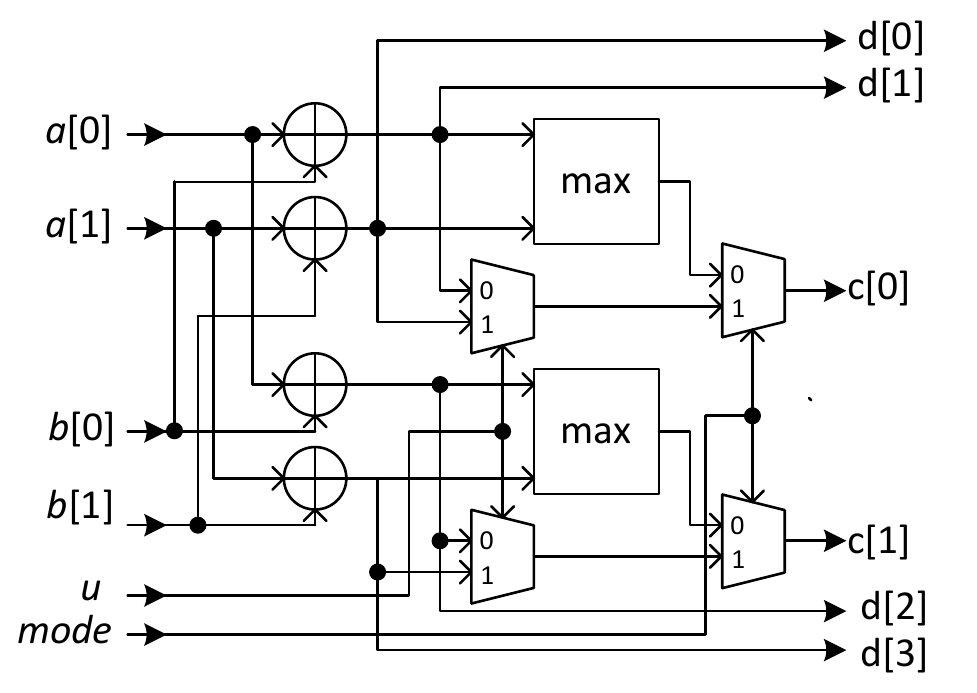}
\caption{Architecture of a processing unit.}
\label{fig:pu_arch}
\end{figure}

The message-passing scheme in MFG of a polar code is in a serial way, which
means that the calculation of a stage depends on the output of its previous
stage. The PUs in \cite{JunPolarList} only carry out the B-TRANS additions. On the
other hand, the S-COMBS stages need only additions and a processing unit has four
adders. Therefore, in order to save hardware resources, the adders in the processing units
is reused to calculate the symbol-based channel transition probability, after these 
processing units finish calculations for the B-TRANS stages. In other words,
additions of both the B-TRANS and the S-COMBS stages are folded onto the
same adders in the processing unites. As shown in Fig.~\ref{fig:pu_arch},
$c[0]$ and $c[1]$ are outputs for the B-TRANS stages; $d[0],d[1],d[2],$ and
$d[3]$ are outputs for the S-COMBS stages. 

\begin{figure}[htbp]
\centering
\includegraphics[width=6cm]{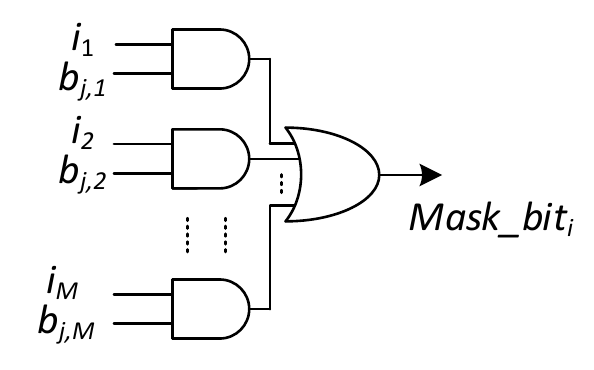}
\caption{Architecture for generating a mask bit.}
\label{fig:MaskBit}
\end{figure}

Block MBG provides a mask bit for each path. 
%For an
%$M$-bit symbol-decision SCL decoder, each path generates $2^{\lvert
%\mathcal{AM}_{j}\rvert}$ paths. 
If there are $f$ $(f geq 0)$ frozen bits in the $M$-bit symbol, the number of
expanded paths will be $2^{M-f}$. For hardware
implementations, we need to consider the worst case and all messages corresponding to
$2^M$ possible paths are calculated. Each path is associated with a mask
bit. When some paths are not needed, due to frozen bits, they are turned off by
mask bits. Fig.~\ref{fig:MaskBit} shows how to
generate the mask bit for path $i$, where $i=(i_1,i_2,\cdots,i_M)\in
\{1,0\}^M$ $(0\leq i
<2^M-1)$ and $\mathbf{b}_j=(b_{j,1},b_{j,2},\cdots,b_{j,M})$ is a frozen-bit indication vector
for $u_{jM+1}^{jM+M}$. If $u_{jM+t}$ is a frozen bit, $b_{j,t}=1$. Otherwise,
$b_{j,t}=0$. If $\mathbf{b}_j$ is an all-one vector, all bits of
$u_{jM+1}^{jM+M}$ are frozen bits, called an $M$-bit frozen vector. If
$\text{\emph{Mask\_bit}}_i$ is 1, $u_{jM+1}^{jM+M}$ is impossible to be
$i$ and the message corresponding to $u_{jM+1}^{jM+M}=i$
is set to 0 in block MSNG.

\begin{figure}[htbp]
\centering
\includegraphics[width=8.5cm]{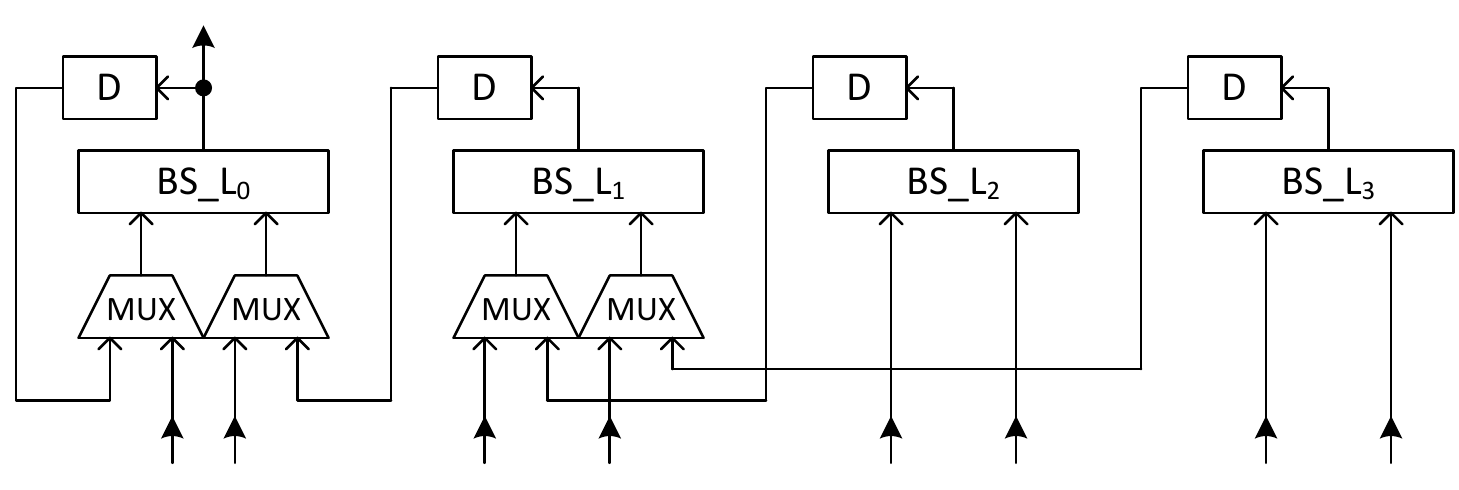}
\caption{Architecture for the folded sorting implementation when $2^M=8$.}
\label{fig:FSN_4}
\end{figure}

\begin{figure}[htbp]
\centering
\includegraphics[width=8.5cm]{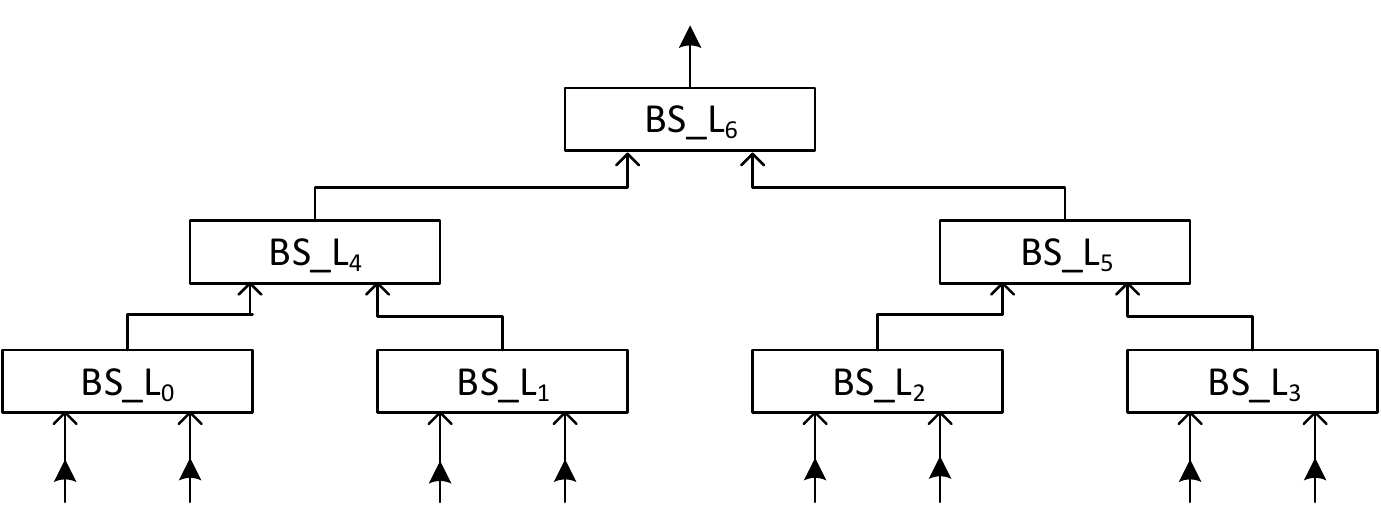}
\caption{Architecture for the tree sorting implementation when $2^M=8$.}
\label{fig:TSN_8}
\end{figure}

Block LPN receives $2^ML$ messages from block MSNG, finds the most reliable $L$
paths, and feeds decision results back to the MPUs. Here, we
use two different sorting implementations -- a
folded sorting implementation and a tree sorting implementation -- for different designs.  The basic unit for
these two implementations is a bitonic sorter \cite{BatcherSorter} ,
which outputs the $L$ max values out of $2L$ inputs. It is referred to as
BS\_L. The folded sorting implementation needs $2^{M-1}$ BS\_Ls $({\rm BS\_L}_0,{\rm
  BS\_L}_1,\cdots,{\rm BS\_L}_{2^{M-1}-1})$. The outputs of the ${\rm BS\_L}_{2i}$
and the ${\rm BS\_L}_{2i+1} (0 \leq i < 2^{M-2})$ are connected with inputs of ${{\rm BS\_L}_i}$
through registers and multiplexers. For the tree sorting implementation with $2^ML$ inputs, $2^M-1$ BS\_Ls are needed. The
tree sorting implementation can be divided into $M$ layers. For $0 \leq i < M$, there
are $2^i$ BS\_Ls in the $i$-th layer. Inputs of the BS\_Ls of the $i$-th layer are connected with
outputs of the BS\_Ls of the $(i+1)$-th layer. Fig.~\ref{fig:FSN_4} and \ref{fig:TSN_8} show examples
of the folded and tree sorting implementations, respectively, for $2^M=8$. 

The folded sorting implementation has a smaller area than the tree sorting
implementation. However, the pipeline can be applied to the tree sorting
implementation by inserting registers between layers to improve the throughput
of the tree sorting implementation.

For the two-stage list pruning network proposed in Sec.~\ref{sec:TSLPN}, either the folded sorting implementation or the tree sorting implementation can be
used for the $2^M$-to-$q$ sorting function and the $qL$-to-$L$ sorting function.

Block CNTL provides control signals to schedule the hardware sharing for
MPUs and decides when to start pruning paths. The signal frz\_flag
is an indicator which is one when a frozen vector appears. When frz\_flag is one, all MPUs use zero to update the
partial-sums instead of outputs of the LPN. In this case, the LPN, the MSNG, and the
calculation of S-COMBS stages are bypassed. The OLG stores
the output paths. The CRCC checks if a path satisfies the CRC constraint. 

%\begin{figure}[htbp]
%\centering
%\includegraphics[width=8cm]{../latex/FSM.pdf}
%\caption{Top architecture for an $M$-bit symbol-decision SCL decoder.}
%\label{fig:SBSCL_arch}
%\end{figure}

%In the following discussion, a simplified semi-parallel SCL polar decoder of the
%list size of $L$ is used as the component decoders. To simplify the calculation of
%messages, the log-likelihood messages are used.
%
%We will focus two aspects in the following text. The first is how to reuse 
%the hardware resources to carry out the calculations of the message-merging block. The second is the design of list
%pruning networks. For different $M$s, we develop different schemes to
%deal with these two aspects. 

\subsection{Message Scheduling and Latency Analysis}
\begin{figure}[htbp]
\centering
\includegraphics[width=8.5cm]{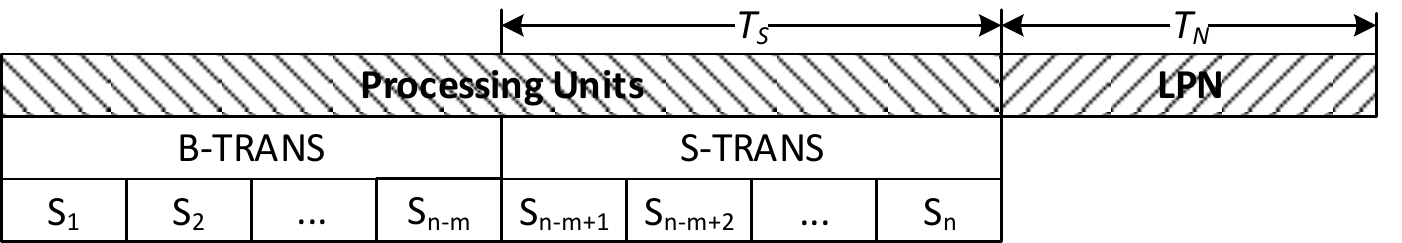}
\caption{Serial scheduling (in clock cycles).}
\label{fig:Ser_SCH}
\end{figure}

To improve area efficiency, for different number of PUs, different scheduling
schemes are needed. To reuse the adders of the processing units, the additions of the S-COMBS stages
in the MFG must be scheduled properly. Assume the number of the processing units is $P$. The total
number of the adders provided by processing units is $4P$. If $2^ML \leq 4P$, we use
a serial scheduling, which means that there is no overlap for the processing units
and the LPN in terms of the operation time, as shown in Fig.~\ref{fig:Ser_SCH}.

Suppose each addition takes one clock cycle. Then each S-COMBS stage takes one
clock cycle to compute messages. Therefore, it takes $m$ clock cycles for the
S-COMBS stages to output messages to the LPN. To save the area, the folded sorting
implementation is applied for the serial scheduling. 

\begin{figure}[htbp]
\centering
\includegraphics[width=8.5cm]{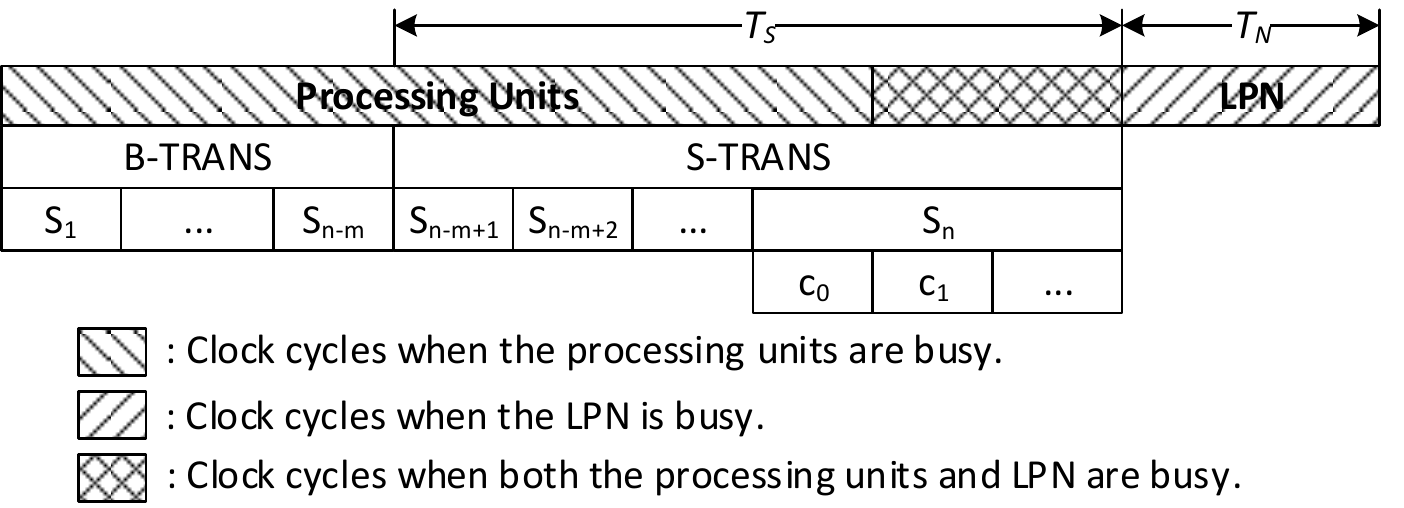}
\caption{Overlapping scheduling (in clock cycles).}
\label{fig:OLP_SCH}
\end{figure}

\begin{figure}[htbp]
\centering
\includegraphics[width=7.5cm]{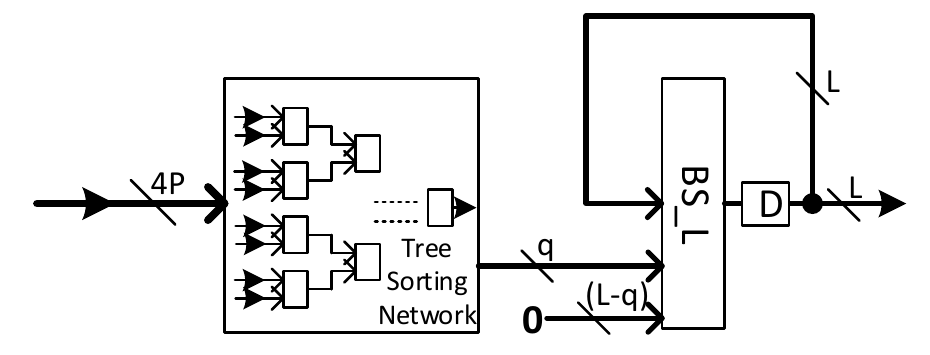}
\caption{A pipelined tree sorting implementation for the overlapping scheduling.}
\label{fig:PP_TREE}
\end{figure}

When $2^ML > 4P \geq 2^{M/2}L$, there are not enough adders to calculate all
$2^ML$ messages of the stage $\text{S}_n$ in one
clock cycle, but all $2^{M/2^{n-i}}L$ messages of the stage $\text{S}_{i}$ $(n+m-1
\leq i \leq n-1)$ can be
calculated in one clock cycle. Without increasing the number of adders, $\frac{2^ML}{4P}$ cycles are needed. In each cycle,
$4P$ messages are calculated. To reduce the latency, the overlapping scheduling
shown in Fig.~\ref{fig:OLP_SCH} is used. In clock cycle ${\rm c}_0$, the first $4P$
messages come out. In clock cycle ${\rm c}_1$, the LPN starts work. Therefore, the MPUs and the
LPN are working simultaneously for $\frac{2^ML}{4P}-1$ clock cycles. Here, the LPN
works in a pipeline way. Hence, the tree sorting implementation is deployed for the
overlapping scheduling and a BS\_L is connected at the end of the tree sorting
implementation in a way shown in Fig.~\ref{fig:PP_TREE}, where the number on a line
represents the number of messages transmitted through the line.

The latency of an $M$-bit symbol-decision SCL decoder consists of: the
latency for calculating messages of the B-TRANS stages, the latency for
calculating messages of the S-COMBS stages, and the latency of the list pruning
network. $T_B$ represents the overall number of clock cycles for the
calculations of the B-TRANS stages. It is equivalent to the latency of a
bit-decision SCL decoder with a code length of $\frac{N}{M}$ and $\frac{P}{M}$
processing units:
\begin{equation*}
T_B=2\frac{N}{M}+\frac{NL/M}{P/M}\log_2(\frac{NL/M}{4P/M})-\frac{NL/M}{P/M},
\end{equation*}
where the third term, $-\frac{NL/M}{P/M}$, is the latency saving
by using PCMS technique. $T_S$ represent the number of clock cycles
for the calculations of S-COMBS stages per symbol. $T_N$ represents the number
of extra clock cycles per symbol needed by the LPN to finish the list pruning after all messages of the stage $S_n$ are
calculated. If
$2^ML \leq 4P$, the number of clock cycles used to calculate messages for
S-COMBS stages is $T_S=m$. When $2^ML > 4P \geq 2^{M/2}L$,
$T_S=m-1+\lceil\frac{2^ML}{4P}\rceil$. More generally, $T_S \leq \sum_{i=1}^{m}\lceil\frac{2^{2^i}L}{4P}\rceil$. $T_N$ is determined by the detailed
implementation. Hence, the latency of the symbol-decision SCL decoder is:
\begin{equation}
\label{eq:SDSCL_LAT}
\begin{split}
T(M)&=(1-\gamma)\frac{N}{M}(T_S+T_N)+T_B \\
&=(1-\gamma)\frac{N}{M}(T_S+T_N)+2\frac{N}{M}+\frac{NL}{P}\log_2(\frac{NL}{8P}),
\end{split}
\end{equation}
where $\gamma$ is a ratio of the number of frozen vectors to $\frac{N}{M}$.

Table \ref{tab:ParLatency} shows the latencies (in clock cycles) for different decoders to decode a
(1024, 480) CRC32-concatenated polar code with 64 processing units and $L=4$. We
assume a BS\_L needs one clock cycle to find the four maximum values out of eight values. For
$M=2$ and $M=4$, a folded sorting implementation and the serial scheduling are
used. For $M=8$, a pipelined tree sorting implementation and the overlapped
scheduling are applied. For $M=8$ and $q=2$, the basic unit in the tree sorting
implementation is to find the two maximum values out of eight values, which needs one
clock cycles. Therefore, $T_N=4$ when $M=8$ and $q=2$.

%\begin{table}[hbtp]
%\begin{center}
%\caption{Latencies for different decoders for a (1024, 480) CRC32-concatenated polar code with 64
%  processing units and $L=4$.}
%\label{tab:ParLatency}
%\begin{tabular}{|c|c|c|c|c|c|}
%\hline
%Decoder & $\gamma$ & $T_R$ & $T_N$ & q & Latency (\# of cycles) \\ \hline
%SDSCL-2 & 0.445 & 1 & 2 & 4 & 2312 \\ \hline
%SDSCL-4 & 0.395 & 2 & 4 & 4 & 1698 \\ \hline
%SDSCL-8 & 0.344 & 6 & 7 & 4 & 1604 \\ \hline
%SDSCL-8 & 0.344 & 6 & 4 & 2 & 1352 \\ \hline
%\end{tabular}
%\end{center}
%\end{table} 

\begin{table}[hbtp]
\begin{center}
\caption{Latencies for different decoders for a (1024, 480) CRC32-concatenated polar code with 64
  processing units and $L=4$.}
\label{tab:ParLatency}
\begin{tabular}{|c|c|c|c|c|c|}
\hline
Decoder & $\gamma$ & $T_S$ & $T_N$ & $q$ & Latency (\# of cycles) \\ \hline
SDSCL-2 & 0.445 & 1 & 2 & 4 & 2069 \\ \hline
SDSCL-4 & 0.395 & 2 & 4 & 4 & 1634 \\ \hline
SDSCL-8 & 0.344 & 6 & 7 & 4 & 1540 \\ \hline
SDSCL-8 & 0.344 & 6 & 4 & 2 & 1288 \\ \hline
\end{tabular}
\end{center}
\end{table} 

It is claimed in \cite{ParSC} that the $M$-bit SDSCL decoder
could have $M$ times faster decoding speed than the bit-decision SCL decoder, which is
much better than our implementation results. Let us
review Eq.~\eqref{eq:SDSCL_LAT} again. For a fair comparison, suppose the MPUs
of the $M$-bit SDSCL decoder has the same architecture as the 
conventional SCL decoder. Then a conventional SCL decoder with the PCMS
technique has a latency of
$T(1)=2N+\frac{NL}{P}\log_2\frac{NL}{8P}$. The decoding speed gain of the
$M$-bit SDSCL decoder is
\begin{equation}
\label{eq:T_ratio}
\begin{split}
\frac{T(1)}{T(M)}&=\frac{2N+\frac{NL}{P}\log_2\frac{NL}{8P}}{(1-\gamma)\frac{N}{M}(T_S+T_N)+2\frac{N}{M}+\frac{NL}{P}\log_2(\frac{NL}{8P})}\\
& = M - \frac{(1-\gamma)N(T_S+T_N)+(M-1)\frac{NL}{P}\log_2\frac{NL}{8P}}{(1-\gamma)\frac{N}{M}(T_S+T_N)+2\frac{N}{M}+\frac{NL}{P}\log_2\frac{NL}{8P}}
\end{split}
\end{equation}

To be exactly $M$ $(M>1)$ times faster,
$(1-\gamma)\frac{N}{M}(T_S+T_N)+(M-1)\frac{NL}{P}\log_2\frac{NL}{8P}$
should be zero. For $NL > 8P$, $\frac{T(1)}{T(M)}<M$, because $T_S\geq 0$ and
$T_N\geq 0$. For $NL = 8P$, $T_S=T_N=0$ should be
satisfied, which means that the calculation of the symbol-based channel
transition probability and the list pruning procedure do \emph{NOT} take any
clock cycle. This is impractical. 
However, $T_S$ and $T_N$
cannot be zero in a practical design. If $NL<8P$ and $P \leq NL$, to achieve $M$ times faster, 
$(T_R+T_N)=\frac{(M-1)L\log_2\frac{8P}{NL}}{(1-\gamma)P}<\frac{5(M-1)}{(1-\gamma)N}$. Usually,
$(1-\gamma)N >> 5(M-1)$. Therefore, the statement about the
decoding speed gain in \cite{ParSC} is too idealistic to be achieved in
practice because the practical implementation needs some extra cycles to calculate the
symbol-based channel transition probability and to perform the list pruning function.

\subsection{Synthesis results}
\label{sec:Syn_results}
To implement the proposed symbol-decision SCL decoder, we consider only $M=2,4$ and
$8$. For $M\geq 16$, it is impractical to build list pruning networks. For
example, for the worst case of $M=16$ that all the bits of a symbol are
information bits, there are $2^{16}L=65536L$ paths. Even if $L=1$, to find the maximum value among $65536$ values still
needs a huge amount of hardware resources and leads to a huge latency.

\begin{table*}[hbtp]
\begin{center}
\begin{threeparttable}[b]
\caption{Synthesis results for different decoders with $L=4$.}
\label{tab:syn_result}
\begin{tabular}{|c|c|c|c|c|c|c||c|c|c|c||c|c|c|c|}
\hline
  & \multicolumn{6}{c||}{Proposed Architectures} &
  \multicolumn{4}{c||}{\cite{ParSC2}}& \cite{JunPolarList}
  &\cite{6823099}\tnote{$\ddagger$} & \cite{6823099}\tnote{$\dagger$} & \cite{LLR-SC-TSP} \\ \hline\hline
Algorithm &\multicolumn{10}{c||}{Symbol-decision SCL} & \multicolumn{4}{c|}{Bit-decision SCL}\\ \hline
$M$ & 2 & 4 & \multicolumn{4}{c||}{8} & \multicolumn{2}{c|}{2} & \multicolumn{2}{c||}{4} & \multicolumn{4}{c|}{N/A} \\ \hline 
%\# of PEs & \multicolumn{6}{c||}{64} & \multicolumn{4}{c||}{4096} & 64 & 64 & 64 & 256 \\ \hline
%$\gamma$ & 0.445 & 0.395 & \multicolumn{4}{c||}{0.344} & \multicolumn{4}{c||}{N/A} & \multicolumn{4}{c|}{0.5} \\ \hline
%$q$ & \multicolumn{4}{c|}{4} &\multicolumn{2}{c||}{2} &
%\multicolumn{8}{c|}{N/A}\\ \hline
Message Type & \multicolumn{13}{c|}{LL} & LLR \\ \hline
Clock Rate (MHz) & \multicolumn{6}{c||}{500} & 525 & 379\tnote{*} & 400 & 289\tnote{*}& 500 & 694 & 314 & 794 \\
\hline
Latency (us) & 4.14 & 3.27 & 3.08 & 3.21\tnote{**}& 2.58 & 2.70\tnote{**} & 3.89 &5.39\tnote{*} &
2.56 &3.53\tnote{*} & 5.63 & 4.06 & 8.25 & 3.34 \\ \hline
Throughput (Mbps) & 247 & 313 & 332 & 319\tnote{**} & 398 & 379\tnote{**} & 262
&189\tnote{*}& 401&289\tnote{*} &182 & 252 & 124 & 307 \\ \hline
Area ($\text{mm}^2$) & 1.126 & 1.209 & 1.669 & 1.782\tnote{**} & 1.403 & 1.519\tnote{**} & 1.98 &
3.79\tnote{*} &
2.14 & 4.10\tnote{*}& 1.099 & 2.197 & 3.53 & 1.78 \\ \hline
Area eff. (Mb/s/$\text{mm}^2$) & 219.4 & 259.2 & 199.2 & 179.1\tnote{**} & 283.3 & 249.3\tnote{**} &
132.3&49.9\tnote{*} & 187.3 &70.6\tnote{*}& 165.6 & 114.7 & 35.1 & 172 \\ \hline
%Normalized Area eff. & 1.29 & 1.51 & 1.15 & 1.08 & 1.63 & 1.51 & 0.30\tnote{**}
%& 0.41\tnote{**} & 1.00 & 1.04 & 0.69 & 0.21 \\ \hline
\end{tabular} 
\begin{tablenotes}
\footnotesize
\item [$\dagger$] The synthesis result in \cite{6823099} is based on a UMC 90nm CMOS technology.
\item [$\ddagger$] The synthesis result is provided by the authors of \cite{6823099}
  based on a TSMC 90nm CMOS technology.
\item [*] Original synthesis results in \cite{ParSC2} are based on an ST 65nm CMOS
  technology. For a fair comparison, synthesis results scaled to a 90nm technology
  are used in the comparison.
\item [**] The design is without the PCMS technique.
\end{tablenotes}
\end{threeparttable}
\end{center}
\end{table*}

In our implementations, $L=4$. Each implementation has 64 processing units. LL
messages are used in our designs. The channel LL messages are quantized with 4 bits. A 
(1024, 480) CRC32-concatenated polar code is used. The synthesis tool is Cadence RTL compiler. The process
technology is TSMC 90nm CMOS technology. Our proposed architectures are compared
with the state-of-the-arts SCL architectures, in \cite{ParSC2, JunPolarList,
  6823099, LLR-SC-TSP}, both bit- and symbol-decision algorithms. The synthesis
results in \cite{LLR-SC-TSP} and \cite{JunPolarList} are also based on a TSMC
90nm CMOS technology. The original synthesis results of \cite{6823099} and
\cite{ParSC2} are based on a UMC 90nm and ST 65nm CMOS technologies, respectively.

%The area efficiency is represented by the
%inverse of the area-latency product and is normalized to that
%of the SCL polar decoder in \cite{JunPolarList}. For the SCL decoder, $\gamma=0.5$ is a ratio of the
%number of frozen bits to the code length $N$.
The synthesis results shown in Table~\ref{tab:syn_result}, demonstrate that our symbol-decision
SCL polar decoders have higher area efficiencies than the SCL decoders in \cite{6823099}, \cite{JunPolarList},
\cite{ParSC2}, and \cite{LLR-SC-TSP}. The SCL decoders in \cite{6823099, LLR-SC-TSP,
  ParSC2} have higher clock rates than our designs because it uses registers as
storage units. However, in our designs, register files are used. 
%Specially, the SDSCL-8 decoder with
%$q=4$ provides a throughput of 319 Mbps and a latency of 3.21 ms. With $q=2$, the SDSCL-8 decoder has a throughput
%of 379 Mbps which is 1.5, 2, and 1.23 times of that of the SCL decoders in
%\cite{6823099}, \cite{JunPolarList} and \cite{LLR-SC-TSP}, respectively. The
%area efficiency of the SDSCL-4 decoder is 2.19, 1.6, 1.45 times of that of the SCL
%decoders in \cite{6823099}, \cite{JunPolarList} and \cite{LLR-SC-TSP}. The area efficiency of our SDSCL-2 (SDSCL-4) decoder is 4.3 (3.51) times
%of that of 2b (4b)-rSCL decoder in \cite{ParSC2}. 

The SDSCL-8 decoders provide a higher throughput and a smaller latency than the
SDSCL-2 and SDSCL-4 decoders, and occupy larger areas. However the improvements
on the throughput and latency are not linear in the symbol size. 

Compared with the SCL decoder in \cite {JunPolarList}, the increase of areas of
symbol-decision SCL decoders is mainly due to sorting networks because the
adders of processing units are reused to calculate both the bit- and
symbol-based channel transition probabilities. For the SDSCL-4 decoder, because the sorting network of
the SDSCL-4 decoder is only 0.073 $\text{mm}^2$, there is no need to shrink $q$
further. For the SDSCL-8 decoders, when $q=4$, the area of the sorting network
is 0.454 $\text{mm}^2$. However, when $q=2$, the sorting network occupies 0.196
$\text{mm}^2$ which is less than a half of that of $q=4$. A smaller $q$ does
help the SDSCL-8 decoder achieve a higher throughput, a smaller latency, a
smaller area, and a higher area efficiency, but it also introduces an FER
performance loss of 0.25 dB to the SDSCL-8 decoder at an FER level of $10^{-3}$
as shown in Fig.~\ref{fig:TSLPP_4}. 
%Compared with the SDSCL-2 and SDSCL-4 decoders, the
%increase of areas of SDSCL-8 decoders is mainly due to the pipelined tree
%sorting network because adders of processing units are reused to calculate the
%symbol-based channel transition probability. 

Moreover, we also provide synthesis results for SDSCL-8 decoders without the
PCMS technique. The PCMS technique helps the SDSCL-8 decoders gain an area
saving of about 0.12 $\text{mm}^2$.

We've already mentioned that LL messages are used in our designs. If LLR
messages \cite{LLR-SC-TSP} are used, symbol-decision SCL decoders can have better area efficiencies than
our current designs because the memory requirement for LLR messages are fewer
than that for LL messages \cite{LLR-SC-TSP}.

\section{Conclusion}
\label{sec:conclusion}
%In this paper, we propose the symbol-based recursive channel transformation for
%symbol-decision polar decoding algorithms. 

In this paper, we use the symbol-based recursive channel combination to
calculate the symbol-based channel transition probability. We show that based on
the LL representation of the transition probability, this recursive procedure
needs fewer additions than the method used in \cite{ParSC, ParSC2}. 
%We prove that the FER of a
% symbol-decision SC algorithm is not worse than that of the
%bit-decision SC algorithm. 
Furthermore, a two-stage list pruning network is proposed
to simplify the $L$-path finding problem. We use the PCMS technique to
reduce the memory requirement for list decoders. By applying the PCMS technique,
we design an efficient architecture for symbol-decision SCL decoders. Specifically, we
introduce two scheduling schemes to perform the hardware sharing. A folded
sorting implementation and tree sorting implementation are also discussed. We also implement
symbol-decision SCL polar decoders for two-bit, four-bit and eight-bit, 
respectively, with a list\ size of four. Our synthesis results show that
symbol-decision SCL polar decoders outperform existing SCL polar decoders in terms of the
area efficiency. Our proposed methods and architecture provide a
range of tradeoffs between area, throughput and area efficiency.

\section*{Acknowledgment}

We would like to thank the authors of \cite{6823099} for providing the
synthesis results using the TSMC 90nm technology in Table~\ref{tab:syn_result}.

%\appendix{Proof of Proposition \ref{prop:1}}
\appendix
\begin{IEEEproof}[Proof of Proposition \ref{prop:1}]
According to the definition of conditional probability $\Pr(B|A)=\frac{\Pr(AB)}{\Pr(A)}$, 
\begin{equation}
\label{eq:prop2}
\begin{split}
W_{2\Lambda,\Phi}^{(i+1)}(y_1^{2\Lambda},&u_1^{i\Phi}|u_{i\Phi+1}^{i\Phi+\Phi})\\
&= \frac{W_{2\Lambda,1}^{(i\Phi+\Phi)}(y_1^{2\Lambda},u_1^{i\Phi+\Phi-1}|u_{i\Phi+\Phi})}{\Pr(u_{i\Phi+1}^{i\Phi+\Phi-1}|u_{i\Phi+\Phi})}.
\end{split}
\end{equation}
Because all bits of $\mathbf{u}$ are independent and each bit has an equal
probability of being a 0 or 1,
\begin{equation*}
\Pr(u_{i\Phi+1}^{i\Phi+\Phi-1}|u_{i\Phi+\Phi}) = \Pr(u_{i\Phi+1}^{i\Phi+\Phi-1})=2^{-(\Phi-1)}.
\end{equation*}
Therefore,
\begin{equation}
\label{eq:prop3}
\begin{split}
W_{2\Lambda,\Phi}^{(i+1)}(y&_1^{2\Lambda},u_1^{i\Phi}|u_{i\Phi+1}^{i\Phi+\Phi})\\
&= 2^{(\Phi-1)}W_{2\Lambda,1}^{(i\Phi+\Phi)}(y_1^{2\Lambda},u_1^{i\Phi+\Phi-1}|u_{i\Phi+\Phi}).
\end{split}
\end{equation}
According to Eq.~\eqref{eq:ArikanT2},
\begin{equation}
\label{eq:prop4}
\begin{split}
W&_{2\Lambda,1}^{(i\Phi+\Phi)}(y_1^{2\Lambda},u_1^{i\Phi+\Phi-1}|u_{i\Phi+\Phi}) \\
&=\frac{1}{2}W_{\Lambda,1}^{(\frac{i\Phi+\Phi}{2})}(y_1^{\Lambda},u_{1,o}^{i\Phi+\Phi-2}\oplus u_{1,e}^{i\Phi+\Phi-2}|u_{i\Phi+\Phi-1}\oplus u_{i\Phi+\Phi})\\
&\cdot W_{\Lambda,1}^{(\frac{i\Phi+\Phi}{2})}(y_{\Lambda+1}^{2\Lambda},u_{1,e}^{i\Phi+\Phi-2}|u_{i\Phi+\Phi}).
\end{split}
\end{equation}

Similarly, we have
\begin{equation}
\label{eq:prop5}
\begin{split}
W&_{\Lambda,1}^{(\frac{i\Phi+\Phi}{2})}(y_1^{\Lambda},u_{1,o}^{i\Phi+\Phi-2}\oplus u_{1,e}^{i\Phi+\Phi-2}|u_{i\Phi+\Phi-1}\oplus u_{i\Phi+\Phi}) \\
&= 2^{-(\frac{\Phi}{2}-1)}W_{\Lambda,\Phi/2}^{(i+1)}(y_1^{\Lambda},u_{1,o}^{i\Phi}\oplus u_{1,e}^{i\Phi}|u_{i\Phi+1,o}^{i\Phi+\Phi}\oplus u_{i\Phi+1,e}^{i\Phi+\Phi}),
\end{split}
\end{equation}
and 
\begin{equation}
\label{eq:prop6}
\begin{split}
W_{\Lambda,1}^{(\frac{i\Phi+\Phi}{2})}(y&_{\Lambda+1}^{2\Lambda},u_{1,e}^{i\Phi+\Phi-2}|u_{i\Phi+\Phi})\\
&= 2^{-(\frac{\Phi}{2}-1)}W_{\Lambda,\Phi/2}^{(i+1)}(y_{\Lambda+1}^{2\Lambda},u_{1,e}^{i\Phi}|u_{i\Phi+1,e}^{i\Phi+\Phi}).
\end{split}
\end{equation}
Then, by equations ~\eqref{eq:prop3} $\sim$ \eqref{eq:prop6},
Eq.~\eqref{eq:prop1} is obtained.
\end{IEEEproof}

% trigger a \newpage just before the given reference
% number - used to balance the columns on the last page
% adjust value as needed - may need to be readjusted if
% the document is modified later
%\IEEEtriggeratref{8}
% The "triggered" command can be changed if desired:
%\IEEEtriggercmd{\enlargethispage{-5in}}

% references section

% can use a bibliography generated by BibTeX as a .bbl file
% BibTeX documentation can be easily obtained at:
% http://www.ctan.org/tex-archive/biblio/bibtex/contrib/doc/
% The IEEEtran BibTeX style support page is at:
% http://www.michaelshell.org/tex/ieeetran/bibtex/
\bibliographystyle{IEEEtran}
% argument is your BibTeX string definitions and bibliography database(s)
%\bibliography{IEEEabrv,../bib/paper}
\bibliography{../latex/bibtex/IEEEfull,../latex/Polar}

%
% <OR> manually copy in the resultant .bbl file
% set second argument of \begin to the number of references
% (used to reserve space for the reference number labels box)
%\begin{thebibliography}{1}
%
%\bibitem{IEEEhowto:kopka}
%H.~Kopka and P.~W. Daly, \emph{A Guide to \LaTeX}, 3rd~ed.\hskip 1em plus
%  0.5em minus 0.4em\relax Harlow, England: Addison-Wesley, 1999.
%
%\end{thebibliography}

% that's all folks
\end{document}